\documentclass[12pt,preprint]{aastex}

\newcommand{\pref}{\protect\ref}

\shorttitle{\indent \def Two components of coronal emission} \shortauthors{Tian et al.}

\begin{document}

\title{Two components of the coronal emission revealed by EUV spectroscopic observations}

\author{Hui Tian\altaffilmark{1}, Scott W. McIntosh\altaffilmark{1}, Bart De Pontieu\altaffilmark{2}, Juan Mart\'{\i}nez-Sykora\altaffilmark{2,3}, Marybeth Sechler\altaffilmark{1}, Xin Wang\altaffilmark{4}}
\altaffiltext{1}{High Altitude Observatory, National Center for Atmospheric Research, P.O. Box 3000, Boulder, CO 80307; htian@ucar.edu}
\altaffiltext{2}{Lockheed Martin Solar and Astrophysics Laboratory, Palo Alto, CA 94304} \altaffiltext{3}{Institute of Theoretical Astrophysics,
University of Oslo, P.O. Box 1029 Blindern, N-0315 Oslo, Norway} \altaffiltext{4}{School of Earth and Space Sciences, Peking University, 100871
Beijing, China}

\begin{abstract}
Recent spectroscopic observations have revealed the ubiquitous presence of blueward asymmetries of emission lines formed in the solar corona
and transition region. These asymmetries are most prominent in loop footpoint regions, where a clear correlation of the asymmetry with
the Doppler shift and line width determined from the single Gaussian fit is found. Such asymmetries suggest at least two emission components: a
primary component accounting for the background emission and a secondary component associated with high-speed upflows. The latter has been proposed to play a vital role in the coronal heating process and there is no agreement on its properties. Here we slightly modify the initially developed technique of Red-Blue (RB) asymmetry analysis and apply it to both artificial spectra and spectra observed by the EUV Imaging Spectrometer onboard Hinode, and demonstrate that the secondary component usually contributes a few percent of the total emission, has a velocity ranging from 50 to 150~km~s$^{-1}$ and a Gaussian width comparable to that of the primary one in loop footpoint regions. The results of the RB asymmetry analysis are then used to guide a double Gaussian fit and we find that the obtained properties of the secondary component are generally consistent with those obtained from the RB asymmetry analysis. Through a comparison of the location, relative intensity, and velocity distribution of the blueward secondary component with the properties of the upward propagating disturbances revealed in simultaneous images from the Atmospheric Imaging Assembly onboard the Solar Dynamics Observatory, we find a clear association of the secondary component with the propagating disturbances.
\end{abstract}


\section{Introduction}
EUV spectroscopic observations often reveal patches of blueshifted emission at boundaries of some active regions
\citep[ARs,][]{Marsch2004,Marsch2008,Harra2008,DelZanna2008,Doschek2008,Tripathi2009,Murray2010,Brooks2011,Warren2011,DelZanna2011}. These authors used a
single Gaussian fit to approximate coronal emission line profiles and derived a blue shift of the order of 20~km~s$^{-1}$. Large line widths
have also been found in these blueshifted regions \citep[e.g.,][]{Doschek2008}. There are suggestions that these blue shifts are indicators of
the nascent slow solar wind \citep{Sakao2007,Harra2008,Doschek2008,Brooks2011}.

Recent investigations have revealed the ubiquitous presence of blueward asymmetries of emission lines formed in the solar corona and transition
region \citep[e.g.,][]{DePontieu2009,McIntosh2009b}. The enhancement in the blue wings of line profiles are most prominent in loop footpoint
regions, such as AR boundaries \citep{Hara2008,DePontieu2009,DePontieu2010,McIntosh2009a,Peter2010,Bryans2010,Tian2011a,Ugarte-Urra2011}. The
discovery of these asymmetric line profiles suggests the presence of a highly blue-shifted (much larger than 20~km~s$^{-1}$) emission component
besides the primary emission component, and thus provides a significant challenge to previous results and interpretations which are based on a single Gaussian
fit.

\cite{DePontieu2009} and \cite{McIntosh2009b} suggested that this secondary emission component is associated with type-II spicules
\citep{DePontieu2007} or rapid blue-shifted events \citep[RBEs,][]{Rouppe2009,DePontieu2011} in the chromosphere, and that they play an
important role in replenishing the corona with hot plasma \citep[e.g.,][]{DePontieu2009,McIntosh2009b,Hansteen2010,DePontieu2011}. The so-called RB (Red-Blue) asymmetry analysis
\citep{DePontieu2009,Martnez-Sykora2011} is based on a comparison of the two wings at same velocity ranges, and detailed analysis indicates that
both the speed (50-150~km~s$^{-1}$) and the relative intensity (a few percent of the total emission) of the secondary component are roughly
consistent across a  temperature range from 100,000 to several million degrees \citep{DePontieu2009,McIntosh2009b}. Through joint imaging and
spectroscopic observations of the corona, \cite{McIntosh2009a} and \cite{Tian2011a} have suggested that the secondary emission component is
caused by high-speed repetitive upflows in the form of upward propagating disturbances in EUV and X-ray imaging observations. Such disturbances
were previously interpreted as slow mode magnetoacoustic waves \citep[e.g.,][]{DeMoortel2000,DeMoortel2002,Robbrecht2001,King2003,McEwan2006,Marsh2009,Wang2009a,Wang2009b,DeMoortel2009,Stenborg2011,Marsh2011} or
slow-speed solar wind outflows \citep[e.g.,][]{Sakao2007,He2010,Guo2010}. There is also a suggestion that they are warps
in two-dimensional sheet-like structures \citep{Judge2011}.

Besides the RB asymmetry analysis, the double Gaussian fit technique has also been used to resolve the second emission component from line
profiles \citep{DePontieu2010,Peter2010,Bryans2010,Tian2011a}. \cite{DePontieu2010} and \cite{Tian2011a} applied a RB guided double Gaussian fit
to spectra with a high signal-to-noise ratio and obvious blueward asymmetry. They used the velocity derived from the RB asymmetry analysis as an
initial guess of the velocity of the secondary component. The algorithm undertakes a global minimization of the difference between the observed
spectrum and the fit by allowing both primary and secondary component centroids to move by one spectral pixel ($\sim$30~km~s$^{-1}$) to the blue or
red of the initial positions. The speed of the secondary component was found to be $\sim$60~km~s$^{-1}$ in \cite{DePontieu2010} and
$\sim$100~km~s$^{-1}$ in \cite{Tian2011a}. They both found that the widths of the two components are similar. The widths of the two Gaussian
components were forced to be the same in the double Gaussian fit algorithm of \cite{Bryans2010}, who mentioned that the velocity of the secondary
component is often as high as 200~km~s$^{-1}$ and that the primary component is also blue-shifted by $\sim$10~km~s$^{-1}$. However, they claimed
that the double Gaussian fit is a good approximation for only the Fe~{\sc{xii}}~195.12\AA{} and Fe~{\sc{xiii}}~202.02\AA{} lines and that other lines are fine with a single Gaussian fit. By applying a
completely free double Gaussian fit to spectral profiles of the Fe~{\sc{xv}}~284\AA{} line, \cite{Peter2010} found that the secondary component
usually contributes 10\% to 20\% to the total emission, is usually blue-shifted by $\sim$40~km~s$^{-1}$ and twice as broad as the primary
component, especially in loop footpoint regions. However, different settings of initial values and allowable ranges of the seven free parameters
can lead to very different fitting results, and thus reasonable initial values and constraints to some free parameters are highly desirable when
performing the double Gaussian fit.

Here we slightly modify the initially developed technique of RB asymmetry analysis and apply it to both artificial and observed spectra by the
EUV Imaging Spectrometer \citep[EIS,][]{Culhane2007} onboard {\it Hinode}, and demonstrate that the speed of the secondary component usually does not
reach 200~km~s$^{-1}$ and that the widths of the two components are comparable. We then use parameters determined from the RB asymmetry analysis
to guide the double Gaussian fit to each spectrum that is observed to have an obvious blueward asymmetry. Both the RB asymmetry analysis and the double
Gaussian fit yield consistent results. We also use imaging observations simultaneously performed by the Atmospheric Imaging Assembly
\citep[AIA,][]{Boerner2010} onboard the Solar Dynamics Observatory ({\it SDO}) to demonstrate that the propagating disturbances at AR edges are plasma upflows which are indeed responsible for the blueward asymmetries of line profiles.

\section{Artificial line profiles}

\begin{figure*}
\centering {\includegraphics[width=0.98\textwidth]{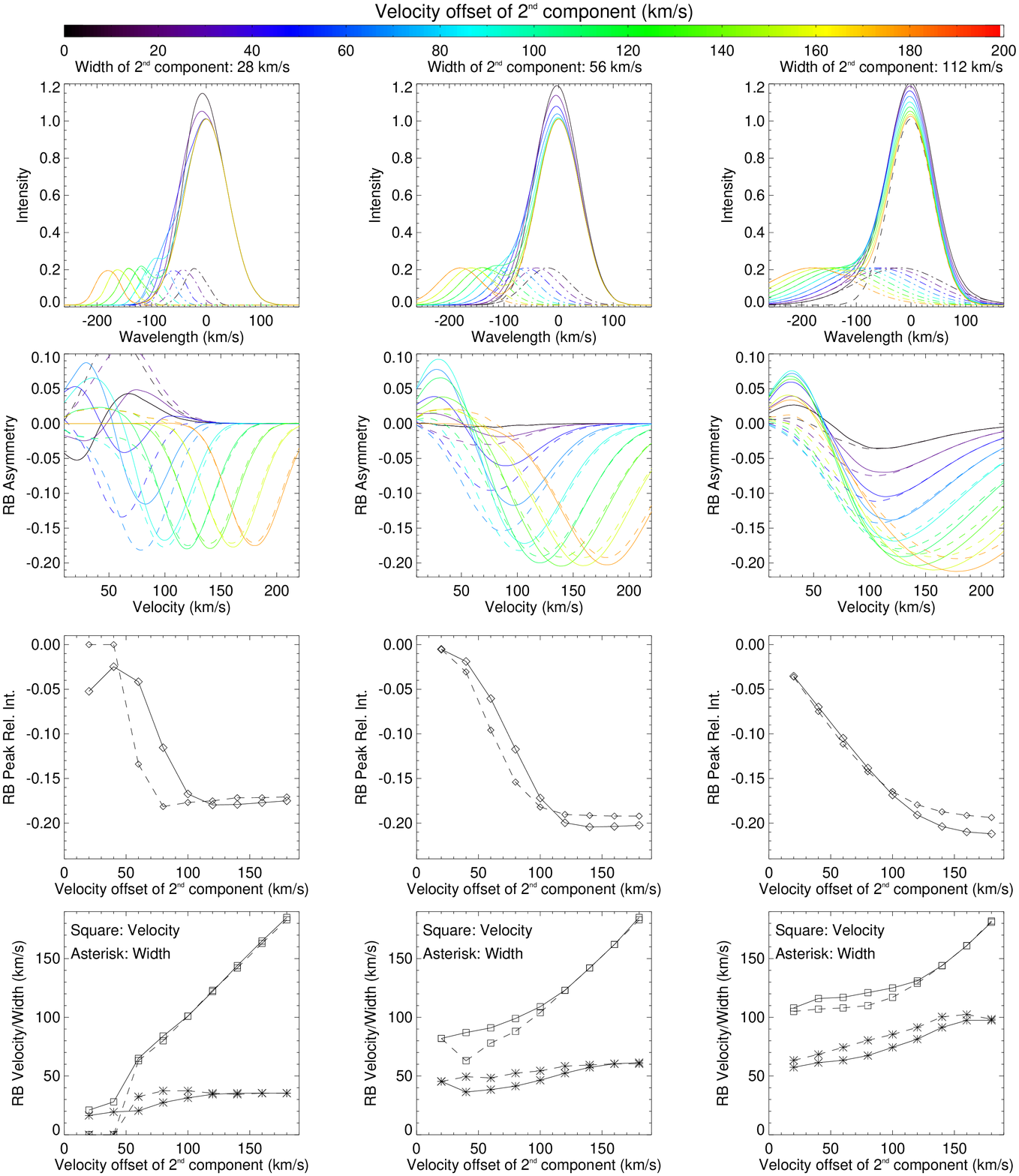}} \caption{Parameters derived from the RB asymmetry profiles of
artificial emission line profiles. First row: the primary, secondary, and total emission profiles are shown as the dashed, dash-dot, and solid
lines, respectively. The peak intensity of the secondary component is set as 20\% of that of the primary one. Second row: RB$_{S}$ (solid) and RB$_{P}$ (dashed) asymmetry profiles. Different colors represent different velocity offsets of the secondary component. Third row: the relative intensity of the peak of the asymmetry profile. Fourth row: the velocity and 1/e width of the asymmetry profile. Solid and dashed lines represent results of RB$_{S}$ and RB$_{P}$, respectively. The Gaussian width of the secondary component is 28, 56, and 112~km~s$^{-1}$ for the left, middle, and right columns, respectively. } \label{fig.1}
\end{figure*}

The technique of RB asymmetry analysis was first introduced by \cite{DePontieu2009}. In our previous work
\citep{DePontieu2009,DePontieu2010,DePontieu2011,McIntosh2009a,McIntosh2009b,McIntosh2010b,Tian2011a,Martnez-Sykora2011}, we first interpolated
the line profile to a spectral resolution ten times greater than the original one, then subtracted the blue wing emission integrated over a
narrow spectral range from that at the same position and over the same range in the red wing. The range of integration was then sequentially
stepped outward from the line centroid to build an RB asymmetry profile (simply RB profile). The RB asymmetry for an offset velocity $u_{c}$ can
be expressed as the following \citep{DePontieu2009,Martnez-Sykora2011}:

\begin{equation}
\emph{$RB(u_c)=\int_{u_c-\delta u/2}^{u_c+\delta u/2}I(u)du-\int_{-u_c-\delta u/2}^{-u_c+\delta u/2}I(u)du$}\label{equation1},
\end{equation}

where $u_{c}$, $\delta u$, and $I(u)$ represent the velocity from the line centroid, the velocity range over which the RB asymmetry is
determined, and the spectral intensity, respectively. The value of $\delta u$ is usually set as 20~km s$^{-1}$. In our previous work, we used the
single Gaussian fit to determine the line centroid. The resulting RB profile can be normalized to the peak intensity derived from the single
Gaussian fit.

Here we slightly modify this technique by using the spectral position corresponding to the peak intensity as the line centroid. The resulting RB
profile is then normalized to the peak intensity of the observed line profile. In the following the RB profiles all refer to normalized RB
profiles. The originally defined and modified techniques are designated as RB$_{S}$ and RB$_{P}$, respectively. In the following we apply these
two techniques to artificial spectra composed of two Gaussian components to test the ability of the RB asymmetry analysis to resolve the
secondary component.

Similar to \cite{Martnez-Sykora2011}, each artificial spectrum consists of a fixed primary component and a much weaker secondary component.
However, to mimic the observed EIS spectra, the Gaussian parameters we used here are very different from those in \cite{Martnez-Sykora2011}. The
spectral pixel size is set as 24 km~s$^{-1}$, similar to that of the EIS spectra at around 274\AA{}. The relative peak intensity of the
secondary component with respect to the primary one can be set as any value below 30\%, while in Figure~\pref{fig.1} we only show the case of
20\%. The Gaussian width (1/e width) of the primary component is set to be 56 km~s$^{-1}$, a value comparable to the coronal line width (including
instrumental, thermal, and non-thermal width) observed by EIS. For the secondary component, three values of the width are chosen (28, 56, 112
km~s$^{-1}$). The spectral position of the secondary component is shifted by -20, -40, -60, ..., -180 km~s$^{-1}$ with respect to the major
component and they are shown as different colors in Figure~\pref{fig.1}.

After applying the RB$_{S}$ and RB$_{P}$ techniques to each artificial spectrum, we obtain RB asymmetry profiles of each spectrum. The same
colors are used in the RB asymmetry plots of Figure~\pref{fig.1}. In the third and fourth rows of Figure~\pref{fig.1} we plot the peak intensity, velocity and 1/e width derived from the RB analysis as a function of the velocity offset of the secondary component.

As noticed by \cite{Martnez-Sykora2011}, when the width of the secondary component is considerably smaller (28 km~s$^{-1}$) than that of the primary one and the secondary component is centered at velocities lower than the Gaussian width of the primary component, the RB$_{S}$ asymmetry profile is blueward at low velocities and redward at higher velocities. The asymmetry profile reverses when the offset velocity of the
secondary component is larger than the width of the primary component. Here we find that the RB$_{P}$ asymmetry profile is redward at low
velocities and blueward at higher velocities when the secondary component is too close to the primary one. As the offset velocity increases to a
value larger than the width of the primary component, the RB$_{P}$ asymmetry profile is blueward at all velocities. We find that when the
velocity offset is larger than the width of the primary component, the velocity, width, and relative intensity derived from the RB$_{P}$
technique are very close to the true velocity and width of the secondary component. The RB$_{S}$ technique does slightly worse but still can
give relatively accurate values of the parameters.

When the widths of two components are the same (56 km~s$^{-1}$), no matter where the secondary component is centered, the RB$_{S}$ asymmetry
profile is redward at low velocities and blueward at higher velocities, and the RB$_{P}$ asymmetry profile is blueward at all velocities. By
comparing the velocity, width, and relative intensity derived from the RB techniques with the true parameters of the secondary component, we can
still conclude that the RB techniques can provide relatively accurate estimate of the properties of the secondary component provided the offset
velocity is larger than the width of the primary component. We notice that compared to the RB$_{S}$ technique, the RB$_{P}$ technique has a
better ability to reproduce the parameters of the secondary component, especially when the offset velocity is smaller than twice the width of
the primary component. The improvement is around 10 km~s$^{-1}$ in the velocity and width, and 4\% in the relative intensity.

When the secondary component is considerably broader (112 km~s$^{-1}$) than that of the primary one, the RB profiles show morphologies similar to
the case of 56 km~s$^{-1}$. However, both the RB$_{S}$ and RB$_{P}$ techniques can not reproduce an offset velocity smaller than 100
km~s$^{-1}$. The velocity derived from RB analysis is always larger than 100 km~s$^{-1}$ if the secondary component is twice as broad as the
primary one. There is also a large discrepancy between the width derived from RB analysis and the true width of the secondary component if the
offset velocity is smaller than twice the width of the primary component (112 km~s$^{-1}$).

We notice that in all the three cases, the RB techniques can give an extremely accurate estimate of the velocity when the offset velocity of the
secondary component is larger than 112 km~s$^{-1}$.

We have varied the intensity ratio of the two components in the range of 5-30\% and obtain basically the same results, except for the different values
of the calculated relative intensities. One thing we have noticed from Figure~\pref{fig.1} is that the RB technique tends to underestimate the value of the relative intensity of the secondary component at smaller offset velocities (e.g., 56-112 km~s$^{-1}$). This is especially the case for the RB$_{S}$ technique which we used in our previous work. From the third row of Figure~\pref{fig.1} we can see that in the velocity range of $\sim$56-112 km~s$^{-1}$ the relative intensity of the secondary component recovered from the RB$_{P}$ technique is much closer to the real value (-0.20 in Figure~\pref{fig.1}), as compared to the RB$_{S}$ technique. The underestimation of relative intensity might be improved by using the Chi-square method \citep{Martnez-Sykora2011}. 

The fact that the calculated RB velocities and relative intensities obviously deviate from the real values when the offset velocities of the secondary component are smaller than the width of the primary component has an important implication for the development of future EUV spectrographs. The EIS instrument has a large instrumental width of $\sim$35 km~s$^{-1}$ \citep[e.g.,][]{Doschek2007}, which contributes almost half of the total line width. The Interface Region Imaging Spectrograph (IRIS), which is scheduled to be launched at the end of 2012, will have a much smaller instrumental width (a Gaussian width of $\sim$5 km~s$^{-1}$ at 1400 \AA{}) and our RB technique should be able to accurately resolve the secondary component at smaller offset velocities. The very high spectral resolution ($\sim$3 km~s$^{-1}$ ) of the IRIS spectrograph will greatly reduce the error caused by the interpolation of line profiles.

In the following we pick out two ARs and investigate properties of the secondary component from the spectra obtained by EIS. Since the modified RB technique, RB$_{P}$, has a better ability to accurately resolve the blueshifted secondary component as compared to the originally defined RB$_{S}$ technique, here we mainly present results by appling the RB$_{P}$ technique instead of the RB$_{S}$ technique to the real data. A comparison between the RB$_{P}$ and RB$_{S}$ results for the 2010 September 16 observation (see the details of this observation in Section 4) is presented in Appendix A.

\section{EIS observation on 2007 January 18}

The $1^{\prime\prime}\times512^{\prime\prime}$ slit of EIS was used to scan AR NOAA 10938 from 18:12 to 20:27 on 2007 January 18, with an
exposure time of 30~s and a step size of  $1^{\prime\prime}$. This data was previously analyzed by \cite{Hara2008} and \cite{Peter2010} for the
purpose of investigating properties of asymmetric line profiles. \cite{Hara2008} mainly used the Fe~{\sc{xiv}}~274.20\AA{} and
Fe~{\sc{xv}}~284.16\AA{} lines which are formed around a temperature of 2~MK, while \cite{Peter2010} only focused on the
Fe~{\sc{xv}}~284.16\AA{} line. Here we choose both lines for our study, but focus on the Fe~{\sc{xiv}}~274.20\AA{} line. \cite{Peter2010}
mentioned that the Fe~{\sc{xv}}~284.16\AA{} line is blended with Al~{\sc{ix}}~284.03\AA{} and that the latter usually contributes no more than
5\% to the deviation from a single Gaussian profile. However, as we will discuss later, in some locations the blend can greatly complicate the
asymmetries of line profiles. The Fe~{\sc{xiv}}~274.20\AA{} line is also a strong line and it is blended with Si~{\sc{vii}}~274.18\AA{}. The
blend is much weaker than Fe~{\sc{xiv}}~274.20\AA{} and can safely be ignored in active region conditions \citep{Young2007}. Moreover, this
blend is very close (less than 1 spectral pixel) to the line center of Fe~{\sc{xiv}}~274.20\AA{} and thus it should not have an important
influence on the results of our RB asymmetry analysis and Gaussian fit. Further confidence is given by the highly similar behavior of the line moments and profile asymmetries between Fe~{\sc{xiv}}~274.20\AA{} and the weaker Fe~{\sc{xiv}}~264.78\AA{} line in most ARs we analyzed (see Appendix B), although the
Fe~{\sc{xiv}}~264.78\AA{} line is usually noisier than Fe~{\sc{xiv}}~274.20\AA{}.

The SSW routine {\it eis\_prep.pro} was applied to correct and calibrate the EIS data. This includes CCD pedestal and dark current subtraction,
cosmic ray removing, warm and hot pixels identification, absolute calibration, error estimation, and so on. The effects of slit tilt and orbital
variation (thermal drift) were estimated by using the SSW routine {\it eis\_wave\_corr.pro} and removed from the data. After that, a running
average over 5 pixels along the slit and 3 pixels across the slit was applied to the spectra to improve the signal to noise ratio.

\begin{figure*}
\centering {\includegraphics[width=0.98\textwidth]{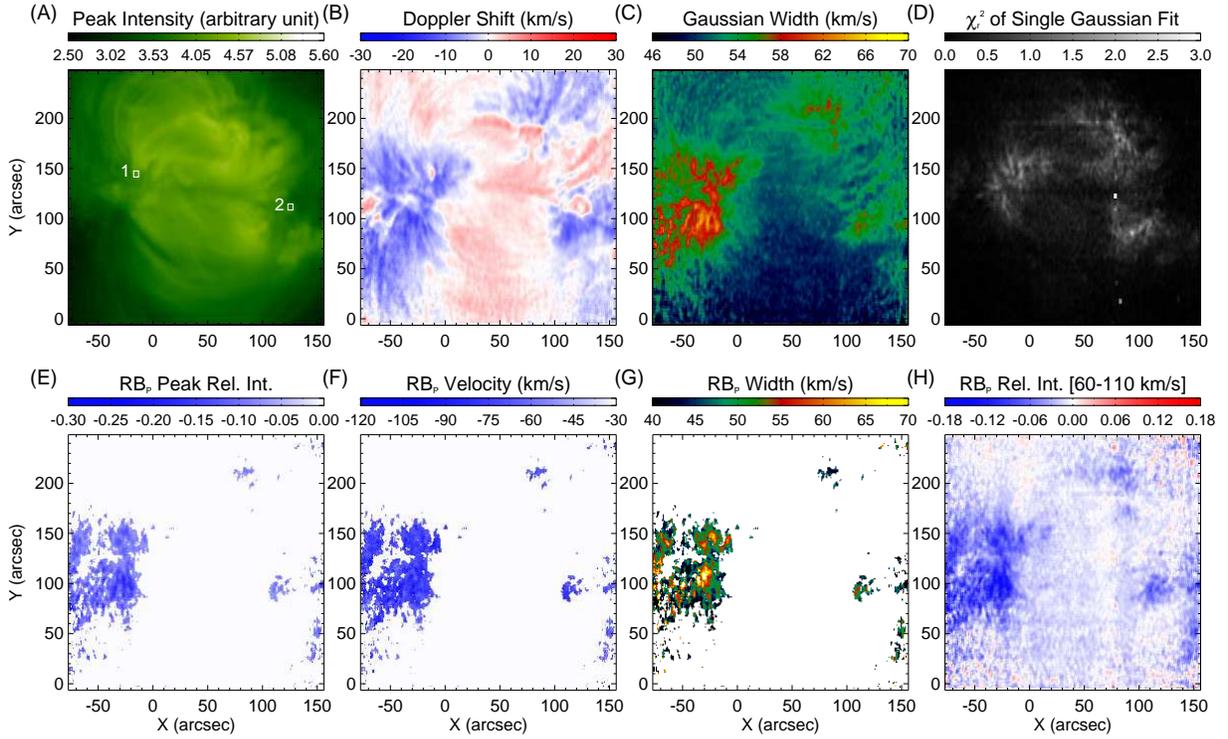}} \caption{Spatial distributions of the parameters derived from the single Gaussian
fit (A-C) and RB asymmetry analysis (RB$_{P}$, E-G) for Fe~{\sc{xiv}}~274.20\AA{}. The map of $\chi_{r}^{2}$ is shown in (D). The map of the
average relative intensity in the velocity interval of 60-110~km~s$^{-1}$, as obtained from the RB profiles, is shown in (H). The two squares in
(A) mark the locations where profiles are averaged and presented in Figure~\pref{fig.4}. } \label{fig.2}
\end{figure*}

As a common practice, a single Gaussian fit was applied to each spectrum and Figure~\pref{fig.2}(A)-(C) show the spatial distributions of the
Gaussian parameters for Fe~{\sc{xiv}}~274.20\AA{}. Figure~\pref{fig.2}(D) shows the map of reduced $\chi_{r}^{2}$ for the fit. The reduced $\chi_{r}^{2}$ is defined as  the following \citep[e.g.,][]{Bevington1992,Peter2001}:

\begin{equation}
\emph{$\chi_{r}^{2}=\sum \frac{1}{N-f}\frac{(d_{i}-m_{i})^{2}}{\sigma_{i}^{2}}$}\label{equation2},
\end{equation}

where $d_{i}$, $m_{i}$, and $\sigma_{i}$ denote the observed spectral radiance, fitted spectral radiance, and measurement error calculated by using the SSW routine {\it eis\_prep.pro} (mainly Poisson error), respectively. Here $i$ represents the spectral position and the summation is performed over all the $N$ spectral positions. The degree of freedom is given by $f$ and equals to 4 or 7 for a single or double Gaussian fit. 

We assume zero shift of the profile averaged over the entire observation region. We can see that the loop footpoint regions, or boundaries of the AR, are characterized by a blueshift of $\sim$20 km~s$^{-1}$ and an enhancement of the line width, a well-known phenomenon in the  Hinode era
\citep[e.g.,][]{Marsch2008,Harra2008,DelZanna2008,Doschek2008,McIntosh2009a,Murray2010,Peter2010,Tian2011a,Brooks2011,Warren2011,DelZanna2011}.
However, the goodness of the fit, the $\chi_{r}^{2}$ is also clearly enhanced mainly in loop footpoint regions, indicating an obvious deviation
from a single Gaussian profile there \citep[e.g.,][]{Peter2001,Peter2010}. Note that the definition of $\chi_{r}^{2}$ in \cite{Peter2010} is not correct and that it is probably a typographic error. 

\subsection{RB asymmetry analysis}

\begin{figure}
\centering {\includegraphics[width=0.48\textwidth]{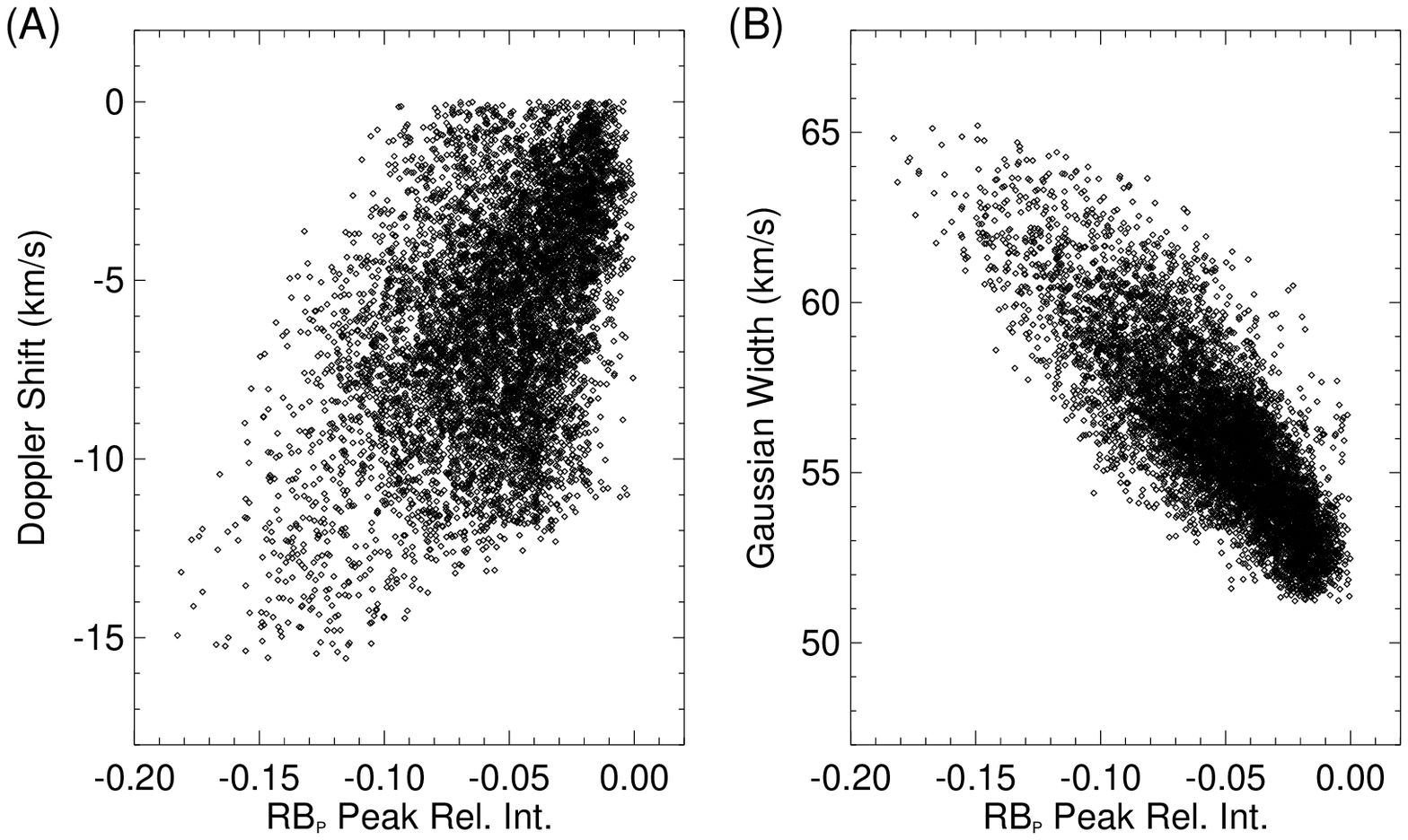}} \caption{Scatter plots of the relationship between Doppler shift/Gaussian width
derived from single Gaussian fit and peak relative intensity derived from RB asymmetry analysis (RB$_{P}$) for Fe~{\sc{xiv}}~274.20\AA{}.}
\label{fig.3}
\end{figure}

We applied the modified RB technique RB$_{P}$ to each spectrum and calculated the average relative intensity in the velocity interval of
60-110~km~s$^{-1}$ from the RB asymmetry profile. Figure~\pref{fig.2}(H) shows the spatial distribution of this average RB asymmetry. Here a
negative/positive value indicates an enhancement of the blue/red wing in this velocity interval. We can see that the blueward asymmetry is most
prominent at loop footpoint regions, generally coincident with the blueshift, width enhancement, and $\chi_{r}^{2}$ enhancement. For further
analysis, we only selected those locations where this average RB asymmetry is smaller than -0.05 (obvious blueward asymmetry) and the signal to noise ratio of the profile (defined as the ratio of the peak and background intensities) is larger than 5. We have to mention that such a cutoff excludes the analysis of spectra in the AR core and edges where the magnetic field lines largely incline with respect to the line of sight, since an inclination of the magnetic field line usually leads to a smaller offset velocity of the secondary component in the line of sight direction, which would reduce the value of the relative intensity derived from the RB technique (see Figure~\pref{fig.1}). 
Figure~\pref{fig.2}(E)-(G) show the peak relative intensity, velocity, and 1/e width derived from RB$_{P}$ asymmetry profiles at these locations. 
The distributions of these three parameters are presented as the red histograms in Figure~\pref{fig.6}(A)-(C).

In Figure~\pref{fig.3} we show the scatter plots of the relationship between Doppler shift/Gaussian width derived from the single Gaussian fit and the 
peak relative intensity derived from RB$_{P}$ asymmetry profiles for Fe~{\sc{xiv}}~274.20\AA{}. Here only data points with blue shift at those
selected locations are shown. A clear correlation is found in each of the two panels in Figure~\pref{fig.3}. Such a correlation strongly suggests
that the clear blue shift and enhanced line width in the loop footpoint regions are related to, or caused by the blueward asymmetries. We can
simply imagine a faint high-speed upflow superimposed on a strong and almost stationary (or slightly shifted) background in the line of sight.
Such a scenario would naturally lead to a blueward asymmetric line profile. And a single Gaussian fit to the total emission line profile would
result in a blueshift and enhanced line width, as compared to the line profile of the background emission-a similar conclusion was reached by \cite{Peter2010}. So it is clear that the blue shift at
AR boundaries, which was reported in many previous investigations
\citep{Marsch2004,Marsch2008,Harra2008,DelZanna2008,Doschek2008,Tripathi2009,Murray2010,Peter2010,Brooks2011,Warren2011,DelZanna2011,He2010}, is actually a composite effect of at least two emission components and can not reflect the real physical process
\citep{McIntosh2009a,DePontieu2010,Peter2010,Bryans2010,Tian2011a}.  Some authors claim that the blue shifts derived from the single Gaussian fit at AR boundaries are indicators of the nascent slow solar wind \citep{Sakao2007,Harra2008,Doschek2008}. If this is the case, the wind speed must have been largely underestimated if the source of the mass supply to the solar wind is the secondary component, or overestimated if the primary component
dominates the mass supply to the solar wind. Our results also suggest that the enhanced line width can be the result of the superposition of different emission components with different velocities and is not necessarily related to the increase in the Alfv\'en wave amplitude \citep{Dolla2011}. We think that both the inhomogeneities of flow velocities and Alfv\'en waves may contribute to the observed non-thermal width of an emission line, and that caution must be taken when interpreting the non-thermal width as purely cased by Alfv\'en waves \citep[e.g.,][]{Banerjee1998,Banerjee2009,Peter2010}. 

From Figure~\pref{fig.6}(A) we can see that the distribution of the intensity ratio (relative intensity of the secondary component) peaks around
0.07 but can reach as high as 0.3. This is generally consistent with previous result that the blueward excess emission is often a few percent of
the total emission, which is based on the RB$_{S}$ technique \citep[e.g.,][]{DePontieu2009,McIntosh2009b}. While \cite{Peter2010} found a
relative intensity of some 10\% to 20\% based on his free double Gaussian fit.

As mentioned above, when the width of the secondary component is not considerably larger than that of the primary one, the RB$_{P}$ technique
can provide an accurate estimate of the Gaussian parameters of the secondary component provided the offset velocity is larger than the width
($\sim$56 km~s$^{-1}$) of the primary component. As seen from Figure~\pref{fig.6}(B), the RB$_{P}$ velocity seldom reaches 100 km~s$^{-1}$ and its
distribution peaks at $\sim$75 km~s$^{-1}$. Such a result would exclude the possibility of a very broad secondary component, as claimed by
\cite{Peter2010} based on a free double Gaussian fit. This is because the velocity derived from RB analysis is always larger than 100 km~s$^{-1}$ when the secondary component is
twice as broad as the primary one (see Figure~\pref{fig.1}). In Figure~\pref{fig.6}(C) we can see that the RB$_{P}$ width is usually smaller than 60
km~s$^{-1}$ and larger than 35 km~s$^{-1}$, and that its distribution peaks around 50 km~s$^{-1}$, which is also not consistent with what we
expect if the secondary component is twice as broad as the primary one. From the last panel of Figure~\pref{fig.1} we can see that the RB$_{P}$ width is always
larger than 60 km~s$^{-1}$ in such a case.  From Figure~\pref{fig.1} we can also see that when the width of the secondary component is
considerably smaller (28 km~s$^{-1}$) than that of the primary one, the RB$_{P}$ width is always smaller than 35 km~s$^{-1}$, which is not
consistent with the observed histogram in Figure~\pref{fig.6}(C). These inconsistencies strongly suggest that the secondary component can not be
considerably broader or narrower than the primary one. The widths of the two components should be comparable. This conclusion implies that the assumption of the same width of the two components in the double Gaussian fit algorithm of \cite{Bryans2010} is reasonable. 

The velocity distribution presented in Figure~\pref{fig.6}(B) is not consistent with the scenario of a very high-speed ($\sim$200 km~s$^{-1}$)
secondary component, as suggested by \cite{Bryans2010} after analyzing another data set. As seen from Figure~\pref{fig.1}, the velocity derived from the RB$_{P}$ technique is
extremely close to (usually less than 5 km~s$^{-1}$) the true velocity of the secondary component when the latter is larger than 112 km~s$^{-1}$.
If very large velocities exist, the RB$_{P}$ technique should yield such large numbers of velocities. However, the observed velocity derived by
using the RB$_{P}$ technique seldom reaches higher than 150 km~s$^{-1}$, indicating that the velocity of the secondary component in our observation can not be as
high as 200 km~s$^{-1}$. We have to mention that the magnitude of velocity also depends on the viewing angle. However, in loop footpoint regions
usually a large portion of the magnetic field lines are almost vertical so that the line of sight effect should not be very significant at disk
center. We also applied the RB asymmetry analysis to the data analyzed by \cite{Bryans2010} and found larger velocities ($\sim$95 km~s$^{-1}$) of the secondary component in that AR (see Appendix C).  But still the velocity seldom reaches 200 km~s$^{-1}$.

\subsection{Double Gaussian fit}

\begin{figure*}
\centering {\includegraphics[width=0.98\textwidth]{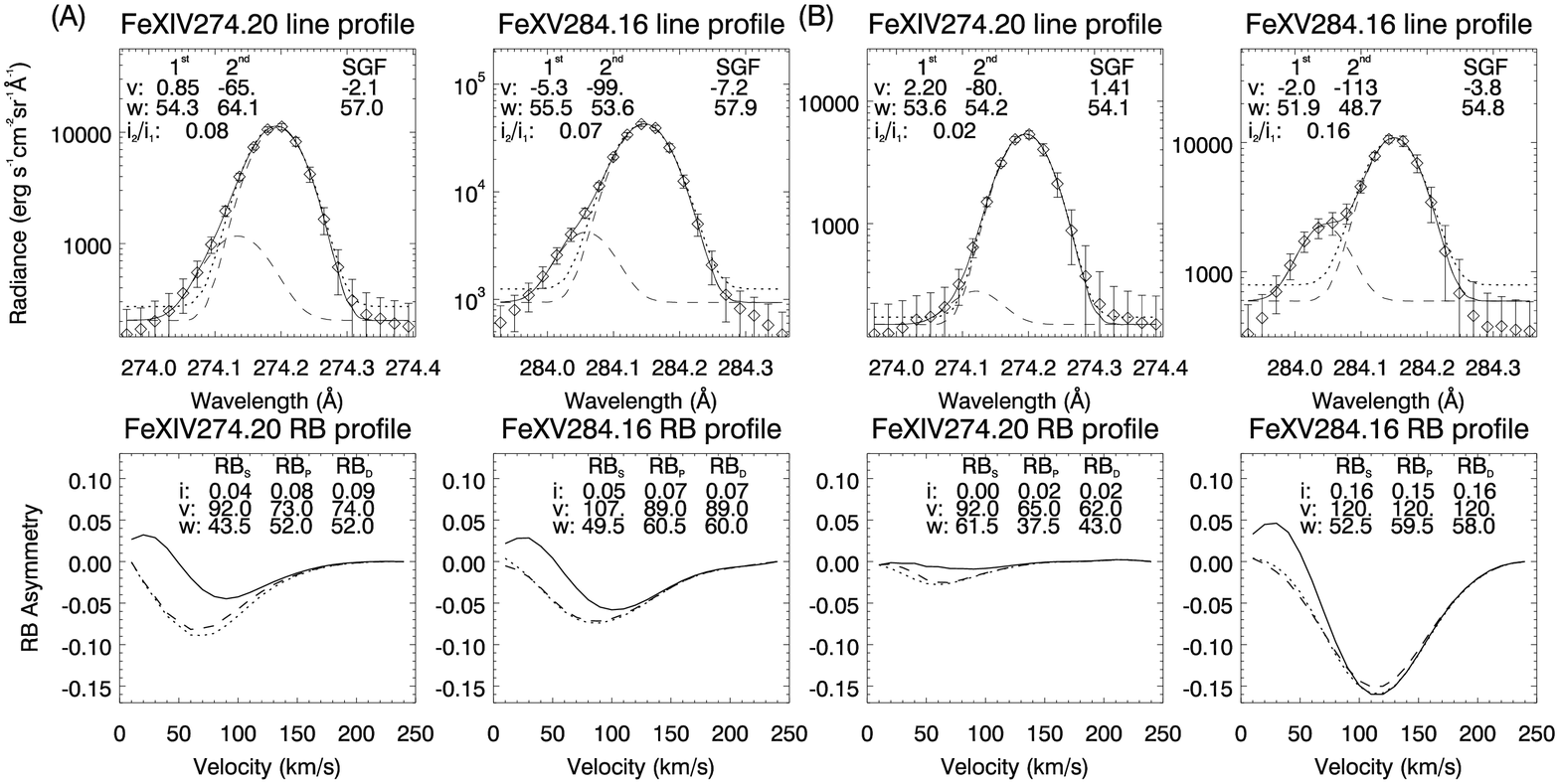}} \caption{ (A) RB asymmetry profiles of the Fe~{\sc{xiv}}~274.20\AA{} and
Fe~{\sc{xv}}~284.16\AA{} line profiles averaged in region 1 marked in Figure~\pref{fig.2}(A). Top: the observed
spectra and measurement errors are shown as the diamonds and error bars, respectively. The dotted lines are single Gaussian fits. The two dashed
lines in each panel represent the two Gaussian components and the solid line is the sum of the two components. The velocity (v) and Gaussian
width (w) derived from the single (SGF) and double (1$^{st}$/2$^{nd}$ for the two components) Gaussian fits are shown in each panel. Also shown
is the intensity ratio of the secondary component to the primary one (i$_{2}$/i$_{1}$). Bottom: the solid, dashed, and dotted lines represent RB
profiles for RB$_{S}$, RB$_{P}$, and RB$_{D}$, respectively. The peak relative intensity (i), velocity (v), and 1/e width (w) are shown in each
panel. (B) Same as (A) but for the line profiles averaged in region 2 marked in Figure~\pref{fig.2}(A).} \label{fig.4}
\end{figure*}

Following the initial idea of using the RB$_{S}$ analysis result to guide double Gaussian fit \citep{DePontieu2010,Tian2011a}, we designed a
similar but modified algorithm. The peak intensity, velocity, and width derived from a single Gaussian fit were used as the initial guess of
those parameters of the primary Gaussian component. The single Gaussian background was also used as the initial value of the double Gaussian
background. Meanwhile, we used the peak intensity and velocity obtained through the RB$_{P}$ analysis as the initial values of the same
parameters of the secondary Gaussian component. The width of the secondary component was initially set as the same as the primary component.
During the iterations, we allowed all the three intensities (peak intensities of the two components and the background intensity) to vary within the range of 75\%-125\% of the corresponding initial values. The
velocity was allowed to move to the blue or red of the initial position by one spectral pixel ($\sim$24~km~s$^{-1}$ at 274\AA{}) for the primary
component and two pixels for the secondary component. We also allowed the width to increase or decrease by one spectral pixel for the primary
component and two pixels for the secondary component. The algorithm undertook a global minimization of the difference between the observed
and fitted spectrum.

We present several examples of the observed and fitted profiles of both the Fe~{\sc{xiv}}~274.20\AA{} and Fe~{\sc{xv}}~284.16\AA{} lines in
Figure~\pref{fig.4}. The profiles in Figure~\pref{fig.4}(A) \& and (B) are the profiles averaged respectively over region 1 and 2 marked in
Figure~\pref{fig.2}. By comparing the observed profiles with the different fitting profiles, we can clearly see the better performance of the
double Gaussian fit and the deviations of the observed profiles from the single Gaussian fits.

After the double Gaussian fit, we took the velocity of the primary component as the line centroid and calculated the RB asymmetry profile (here
RB$_{D}$). The lower panels of Figure~\pref{fig.4} show the three RB asymmetry profiles (RB$_{S}$, RB$_{P}$, RB$_{D}$). We can see that RB$_{P}$
and RB$_{D}$ show similar behaviors and values of the three parameters are also very close.

It is known that the Fe~{\sc{xv}}~284.16\AA{} line is blended with Al~{\sc{ix}}~284.03\AA{} \citep[e.g.,][]{Young2007,Brown2008}. The latter is sitting at $\sim$150~km~s$^{-1}$ away from the spectral position of Fe~{\sc{xv}}~284.16\AA{} so that this blend is potentially contaminating the blue wing of the line and may complicate the RB analysis and Gaussian fit. For example, the line profiles in both Figure~\pref{fig.4}(A) and (B) were obtained around loop footpoint regions (marked in Figure~\pref{fig.2}). The magnitude of asymmetry is comparable in both the Fe~{\sc{xiv}}~274.20\AA{} and Fe~{\sc{xv}}~284.16\AA{} line profiles shown in Figure~\pref{fig.4}(A), indicating that the blend Al~{\sc{ix}}~284.03\AA{} can be ignored. However, in Figure~\pref{fig.4}(B) we see a $\sim$2\% blueward asymmetry in Fe~{\sc{xiv}}~274.20\AA{} but a $\sim$16\% blueward asymmetry in Fe~{\sc{xv}}~284.16\AA{}. Since the two lines have similar formation temperatures, the completely different magnitude of asymmetries in the two lines strongly suggests that the bump at the blue wing of Fe~{\sc{xv}}~284.16\AA{} is dominated by the blend Al~{\sc{ix}}~284.03\AA{}. Therefore, we focus on the Fe~{\sc{xiv}}~274.20\AA{} line instead of Fe~{\sc{xv}}~284.16\AA{} which was used by \cite{Peter2010}.

\begin{figure*}
\centering {\includegraphics[width=0.98\textwidth]{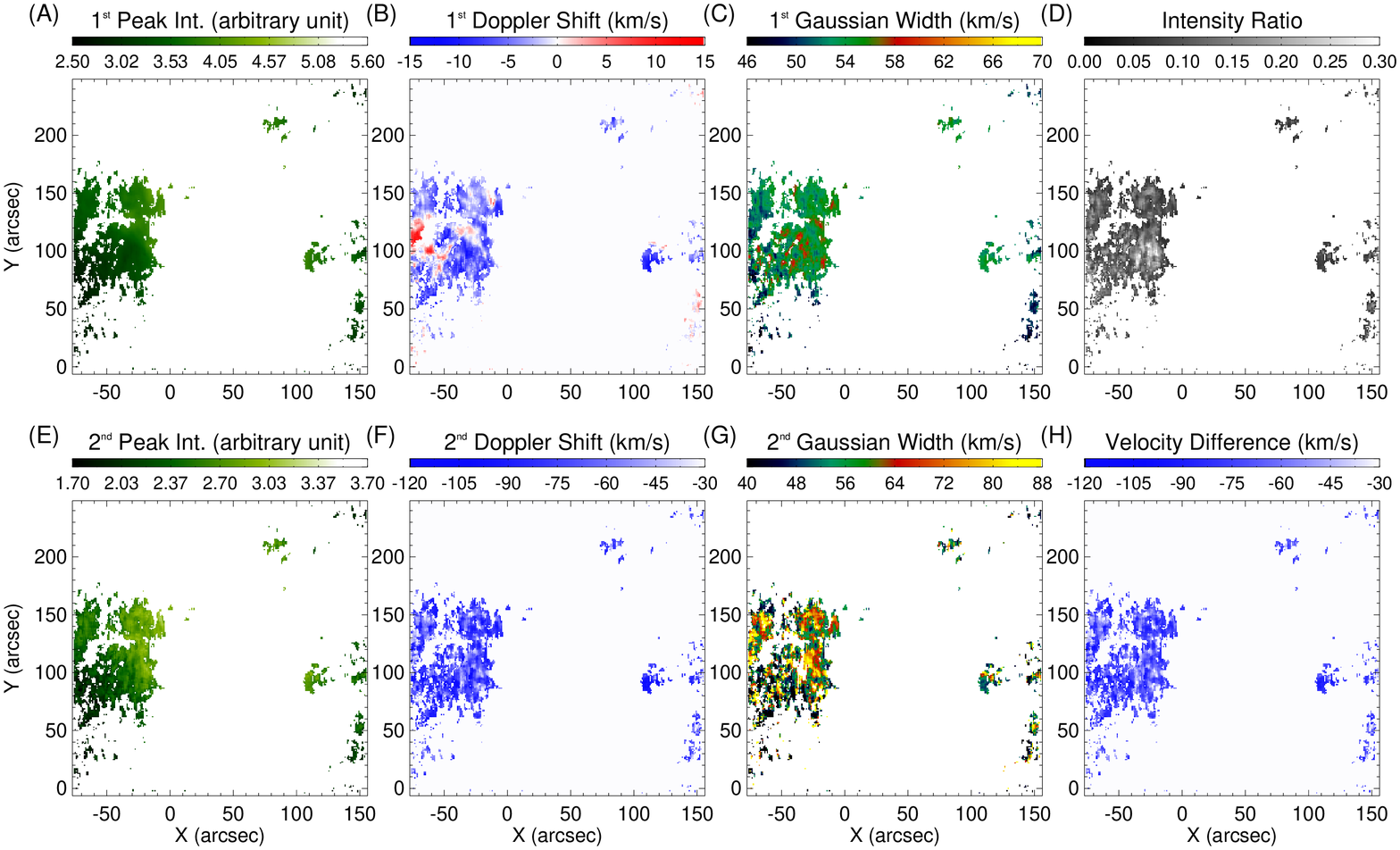}} \caption{Parameters of the primary (A-C) and secondary (E-G) components as derived
from the double Gaussian fit for Fe~{\sc{xiv}}~274.20\AA{}. The intensity ratio and velocity difference of the two components are shown in
(D) and (H), respectively. } \label{fig.5}
\end{figure*}

We then applied the RB$_{P}$ guided Gaussian fit and RB$_{D}$ analysis to all spectra of Fe~{\sc{xiv}}~274.20\AA{} with obvious blueward asymmetries and large signal to noise ratio. We present the spatial
distributions of the three Gaussian parameters of both components in Figure~\pref{fig.5}. In addition, the intensity ratio and velocity difference
of the two components are also shown in Figure~\pref{fig.5}. We found that in most locations the primary component is slightly blue shifted
($\sim$10~km~s$^{-1}$), a result also found by \cite{Bryans2010}, who interpreted it as the nascent slow solar wind from AR boundaries \citep{Wang2009}. While for the secondary component, the velocity is largely blue shifted
($\sim$80~km~s$^{-1}$) at almost all locations. The maps of parameters derived from RB$_{D}$ asymmetry profiles are very similar to those
derived from RB$_{P}$ and thus are not shown here. 

\begin{figure}
\centering {\includegraphics[width=0.48\textwidth]{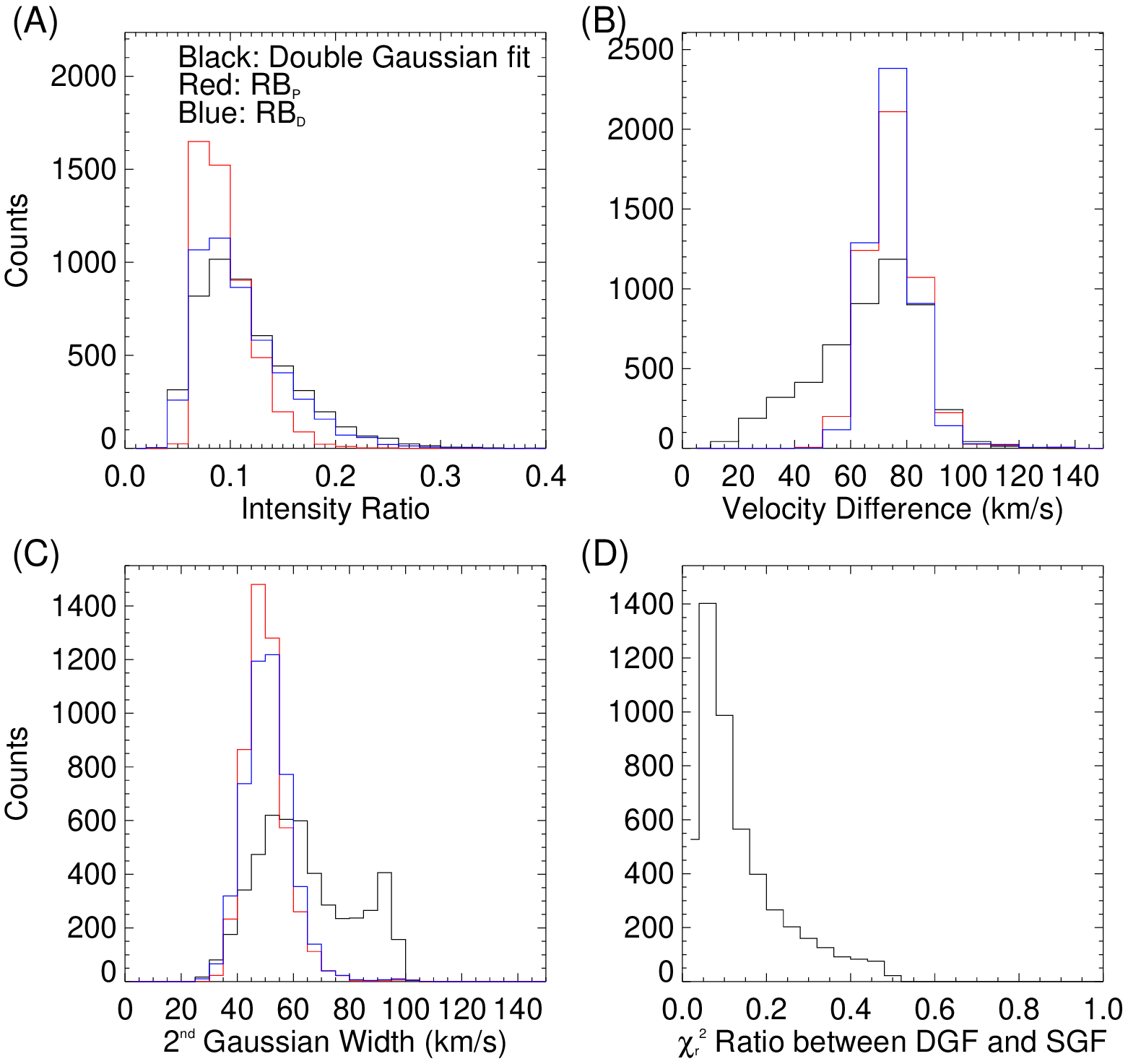}} \caption{ Histograms of the intensity ratio (A), velocity difference (B), and
Gaussian width of the secondary component (C), as derived from double Gaussian fit (black) and RB asymmetry analysis (red/blue for
RB$_{P}$/RB$_{D}$) for Fe~{\sc{xiv}}~274.20\AA{}. Panel (D) shows the histograms of the $\chi_{r}^{2}$ ratio between the double and single
Gaussian fit. } \label{fig.6}
\end{figure}

Histograms of the intensity ratio (relative intensity), velocity difference (velocity), and Gaussian width of the secondary component as derived
from double Gaussian fit and RB asymmetry analysis (RB$_{P}$,RB$_{D}$) are presented in Figure~\pref{fig.6}. We can see that they are very
similar, especially RB$_{P}$ and RB$_{D}$. The intensity ratio peaks around 0.07 and can also reach as high as 0.3. For RB$_{P}$ and RB$_{D}$,
the velocity and width both have a narrow distribution which peaks at $\sim$75~km~s$^{-1}$ and $\sim$50~km~s$^{-1}$, respectively. While for the
double Gaussian fit, most of the width values are still much smaller than $\sim$100~km~s$^{-1}$ and the width distribution peaks slightly
higher. The velocity difference is usually in the range of 50-100~km~s$^{-1}$, but the double Gaussian fit algorithm also produces some
relatively small values of velocity difference (10-50~km~s$^{-1}$). However, these relatively small blue shifts are found at only a few
locations and they are not prevalent.

The $\chi_{r}^{2}$ value of the double Gaussian fit is usually smaller than unity-similar result was also found by \cite{Peter2010}, who attributed it to the overestimation of the measurement error. From Figure~\pref{fig.6}(D) we can see that the $\chi_{r}^{2}$ ratio between the double and single Gaussian fit is usually less than 0.5 in the regions characteristic of obvious blueward asymmetries. Such a result indicates that the double Gaussian fit does at least two times better than the single Gaussian fit for these asymmetric line profiles. It is likely that even more components, e.g., a component from cooling downflows, might be present in the observed emission. However, a reliable decomposition of these additional components can not be made due to the relatively large instrumental width and modest spectral resolution of EIS. With the upcoming IRIS data, we may be able to resolve more coronal emission components.

\section{EIS and AIA observations on 2010 September 16}

The other observation was performed by EIS on 2010 September 16, with simultaneous observations by AIA. The EIS instrument scanned NOAA AR
11106 from 10:38 to 11:57 UT. The $2^{\prime\prime}\times512^{\prime\prime}$ slit was used for the observation, with a 60~s exposure. After standard correction and calibration of the EIS data, a running average over 5 pixels along the slit and 3 pixels across the slit was applied to the spectra to improve the signal to noise ratio. Four relatively strong and clean emission lines in the spectral window were selected for our study: Si~{\sc{vii}}~275.35\AA{}, Si~{\sc{x}}~258.37\AA{}, Fe~{\sc{xiii}}~202.04\AA{}, and Fe~{\sc{xiv}}~274.20\AA{}. Their formation temperatures are log({\it T}/K)=5.8, 6.1, 6.2, and 6.3, respectively \citep{Young2007}. But we will mainly concentrate on the hottest strong line, Fe~{\sc{xiv}}~274.20\AA{}.

\begin{figure*}
\centering {\includegraphics[width=0.98\textwidth]{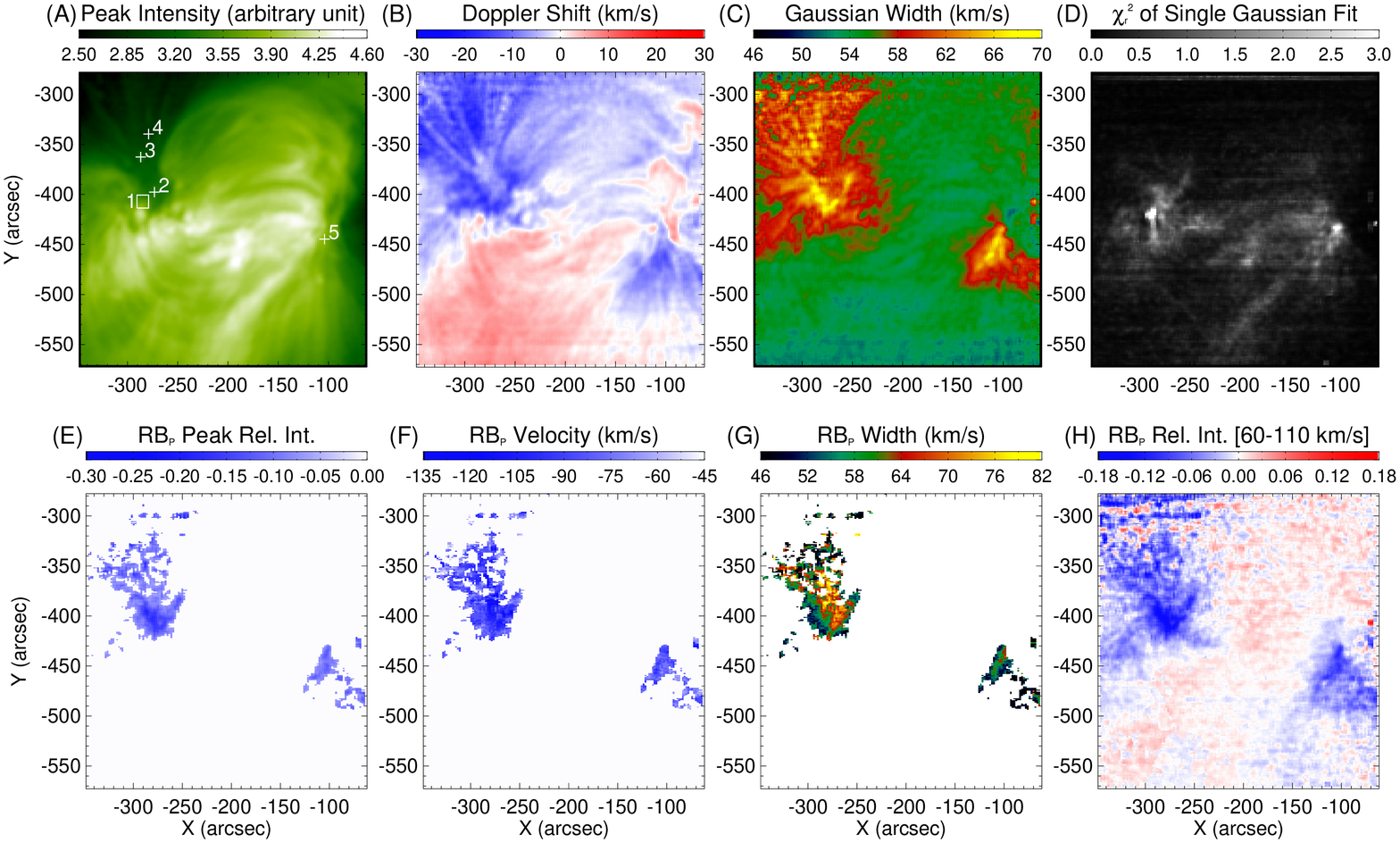}} \caption{Same as Figure~\pref{fig.2} but for the observation on 2010 September 16. The
square in (A) mark the location where profiles of several emission lines are averaged and presented in Figure~\ref{fig.9}(A). The four pluses mark
locations of the Fe~{\sc{xiv}}~274.20\AA{} line profiles presented in Figure~\ref{fig.9}(B)-(E)} \label{fig.7}
\end{figure*}

A single Gaussian fit was first applied to each spectrum to derive the line intensity, centroid and Gaussian width. By assuming the average Doppler shift of each line is zero over the entire region, we calculated the Doppler shift from the line centroid for each spectral profile. Maps of the Gaussian parameters and $\chi_{r}^{2}$ for Fe~{\sc{xiv}}~274.20\AA{} are presented in Figure~\pref{fig.7}(A)-(D). Again, we clearly see that the loop footpoint regions, or boundaries of the AR, are characterized by a blueshift of $\sim$20 km~s$^{-1}$ and an enhancement of the line width. The deviation from a single Gaussian profile, which is quantified by the enhancement of $\chi_{r}^{2}$, is also clearly seen in the loop footpoint regions. The Si~{\sc{x}}~258.37\AA{} and Fe~{\sc{xiii}}~202.04\AA{} lines reveal basically the same structures in the maps of Gaussian parameters. The cool line Si~{\sc{vii}}~275.35\AA{} shows prominent redshifts in the fan regions (not shown here), a phenomenon found also by \cite{Warren2011}. Such redshifts could be the downflows after cooling of the high-speed upflows and evidence for the mass circulation in coronal loops \citep{Marsch2008}. 

\subsection{RB asymmetry analysis}

\begin{figure}
\centering {\includegraphics[width=0.48\textwidth]{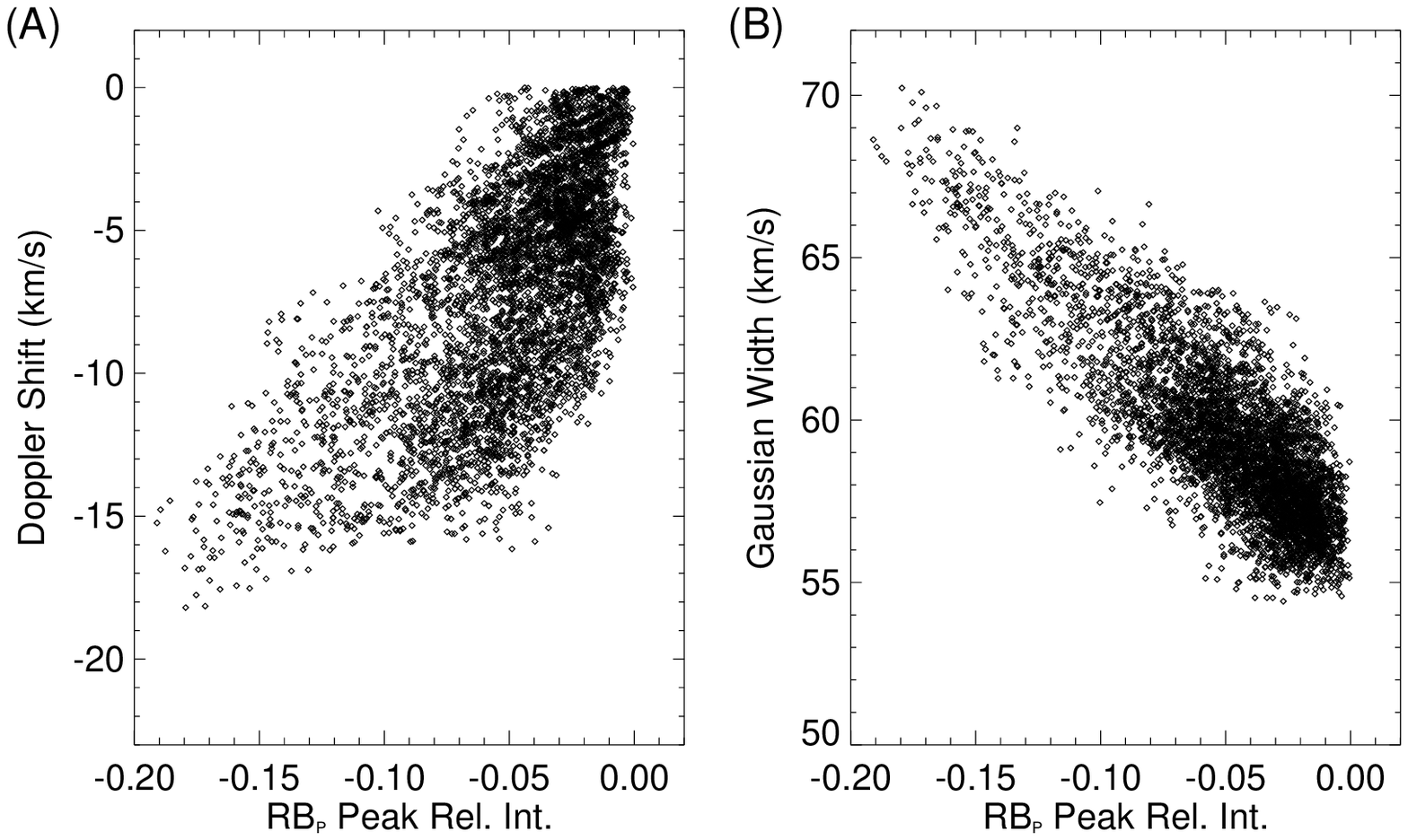}} \caption{Same as Figure~\pref{fig.3} but for the observation on 2010 September 16.}
\label{fig.8}
\end{figure}

Similarly to before, we applied the RB$_{P}$ technique to each spectrum and calculated the average relative intensity in the velocity interval of
60-110~km~s$^{-1}$ from the RB asymmetry profile. Figure~\pref{fig.7}(H) shows the spatial distribution of this average RB asymmetry. It is clear that the blueward asymmetry is most prominent at the two edges of the AR, generally coincident with the enhancement of blueshift, line width, and $\chi_{r}^{2}$. However, the blueward asymmetry is clearly not only present at the AR edges, but also existing in other plage/loop footpoint regions. For example, the tongue-like patch (x=$-230^{\prime\prime}$,y=$-430^{\prime\prime}$) of blueward asymmetries in Figure~\pref{fig.7}(H) seems to be coincident with moss and it is not a typical outflow region. The fact is that simple magnetic field structures which are aligned with the line of sight are often not present in AR cores so that the secondary component there is often not revealed as clear blueward asymmetries of line profiles. An obvious correlation between the Doppler shift/Gaussian width derived from single Gaussian fit and peak relative intensity derived from RB$_{P}$ asymmetry profiles is found in Figure~\pref{fig.8}, suggesting that the blueshift and enhanced line width are caused by the high-speed secondary emission component and that the single Gaussian fit can not accurately reflect the real physical process here. In the following, only those pixels with obvious blueward asymmetry (the average RB asymmetry in 60-110~km~s$^{-1}$ smaller than -0.05) and significant signal to noise ratio (larger than 5) were selected. Clearly, such a cutoff excludes the analysis of spectra in the AR core and edges where the magnetic field lines make large angles relative to the line of sight. The peak relative intensity, velocity, and 1/e width derived from RB$_{P}$ asymmetry profiles of Fe~{\sc{xiv}}~274.20\AA{} at these pixels are shown in Figure~\pref{fig.7}(E)-(G). 

The distributions of these three parameters are shown as the red histograms in Figure~\pref{fig.11}(A)-(C) and they are very similar to those in Figure~\pref{fig.6}. The only obvious difference is perhaps the slightly larger peak velocity and Gaussian width. In the 2007 January 18 observation, the velocity and width peak at $\sim$75~km~s$^{-1}$ and $\sim$50~km~s$^{-1}$, respectively. But in the 2010 September 16 observation, the velocity peaks at $\sim$85~km~s$^{-1}$ and the width peaks at $\sim$60~km~s$^{-1}$. The major part of the velocity distribution is still on the left side of 100 km~s$^{-1}$ in the 2010 September 16 observation, which is not consistent with the scenario of a considerably broader secondary component \citep{Peter2010}. The scenario of a considerably narrow secondary component can also be excluded since the RB$_{P}$ width is almost always larger than 35 km~s$^{-1}$. Similar to the 2007 January 18 observation, the RB$_{P}$ velocity in this recent observation also seldom reaches higher than 150 km~s$^{-1}$, indicating that the line of sight velocity of the secondary component is unlikely to be as high as 200 km~s$^{-1}$ in this observation.
So with the RB asymmetry analysis (particularly RB$_{P}$), we can conclude that the secondary emission component, which is usually less than 20\% and often a few percent of the primary one in intensity, has a speed around $\sim$80~km~s$^{-1}$ and a width comparable to the primary component. 

\subsection{Double Gaussian fit}

\begin{figure*}
\centering {\includegraphics[width=0.98\textwidth]{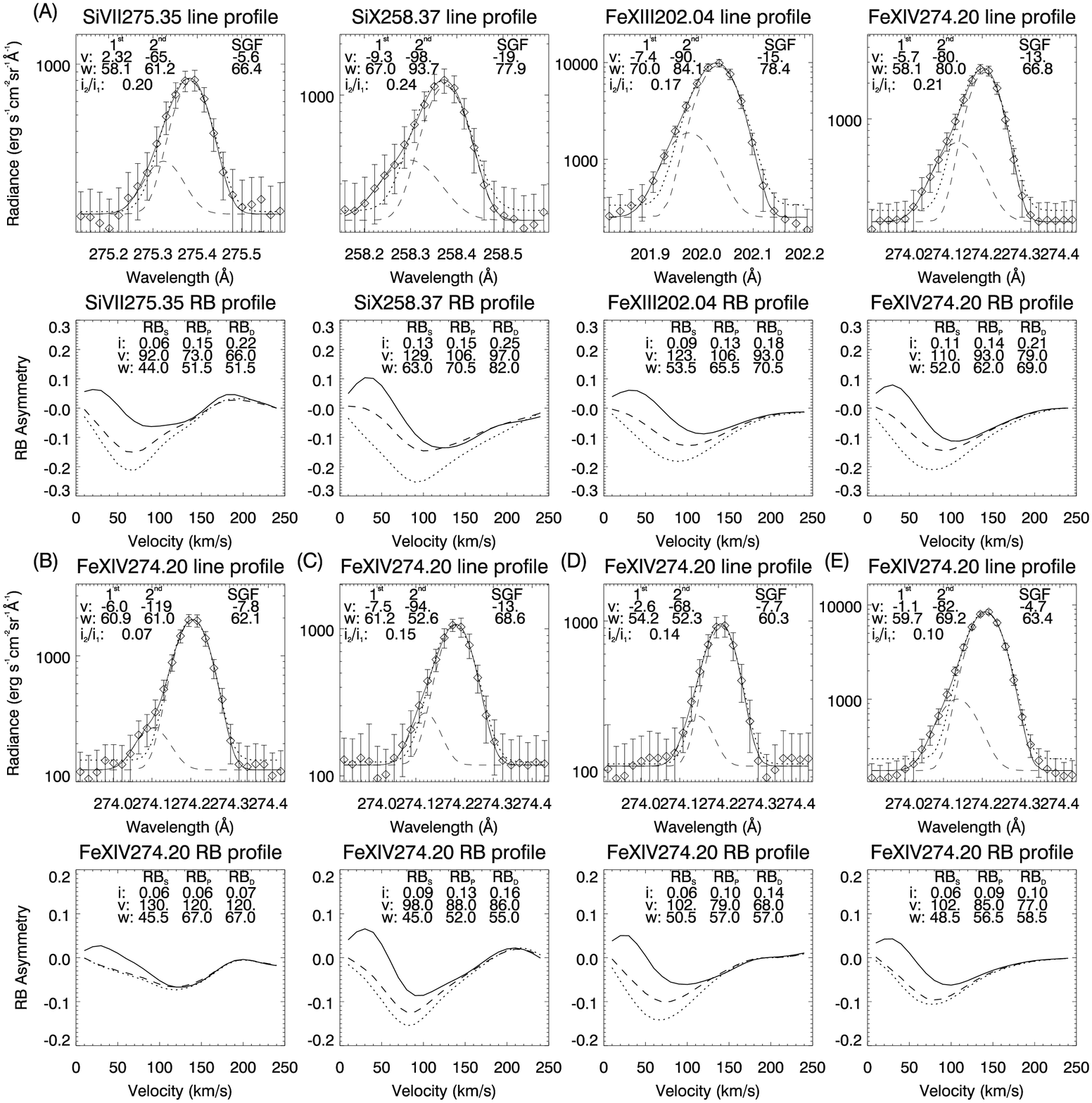}} \caption{(A) RB asymmetry profiles (bottom) of four emission line profiles (top)
averaged in the square marked in Figure~\pref{fig.7}(A). (B)-(E) RB asymmetry profiles (bottom) of the Fe~{\sc{xiv}}~274.20\AA{} line profiles
(top) at locations 2-5 marked in Figure~\pref{fig.7}(A). The line styles and denotations of parameters are the same as in Figure~\pref{fig.4}.}
\label{fig.9}
\end{figure*}

The same algorithm of RB$_{P}$ guided double Gaussian fit was adopted to the asymmetric profiles obtained in the 2010 September 16 observation.
Several examples of the observed and fitted line profiles are presented in Figure~\pref{fig.9}. In Figure~\pref{fig.9}(A) we show profiles of four emission lines averaged over region 1 which is marked in Figure~\pref{fig.7}. Blueward asymmetries and deviations from single Gaussian profiles are clearly present in all of the four lines. The double Gaussian algorithm yields a reasonably good fit to the observed spectra. This spatial average is necessary because individual profiles of the Si~{\sc{vii}}~275.35\AA{} and Si~{\sc{x}}~258.37\AA{} lines are too noisy to allow reliable Gaussian fits and RB asymmetry analysis in the AR boundaries. 

We find that the relative intensity and velocity of the secondary component stay relatively stable across the temperature range of log({\it T}/K)=5.8-6.3, consistent with the scenario of multi-thermal high-speed upflows, as suggested by our previous work \citep{DePontieu2009,DePontieu2010,DePontieu2011,McIntosh2009a,McIntosh2009b,McIntosh2010a,McIntosh2010b,Tian2011a,Tian2011b,Martnez-Sykora2011}. The clear presence of blueward asymmetry of the cool Si~{\sc{vii}}~275.35\AA{} line is consistent with the finding of \cite{McIntosh2009a}. Compared to other lines, the Si~{\sc{vii}}~275.35\AA{} line reveals a smaller velocity ($\sim$65~km~s$^{-1}$) of the secondary component. One possible reason could be that the Si~{\sc{vii}}~275.35\AA{} line reflects emission from much cooler plasmas so that its line profiles are complicated by cooling downflows \citep[see also][]{Ugarte-Urra2011}. 

Fe~{\sc{xiv}}~274.20\AA{} line profiles at four single pixels (without spatial average) are presented in Figure~\pref{fig.9}(B)-(E). The secondary component resolved by the double Gaussian fit has different relative intensity and velocity. The magnitude of velocity is likely to be related to the viewing angle of the upflows. If the magnetic field lines associated with the upflows make a smaller angle with respect to the line of sight, the observed velocity should be closer to the real velocity. The location of the profile presented in Figure~\pref{fig.9}(B) is in a weak emission region surrounded by hot loops. This weak emission region might be related to less compact (larger) loops or open field lines which are usually more vertical at lower heights compared to compact and smaller loops. This may explain the large value of velocity ($\sim$120~km~s$^{-1}$). The Gaussian widths of the two components are generally comparable in all these four cases. 

The RB$_{D}$ profiles were then calculated by taking the velocity of the primary component as the line centroid. In Figure~\pref{fig.9} we also show the three RB asymmetry profiles (RB$_{S}$, RB$_{P}$, RB$_{D}$) for the corresponding spectra. The parameters of the secondary component obtained through RB$_{P}$ and RB$_{D}$ techniques are usually close to those obtained by using the double Gaussian fit. 

\begin{figure*}
\centering {\includegraphics[width=0.98\textwidth]{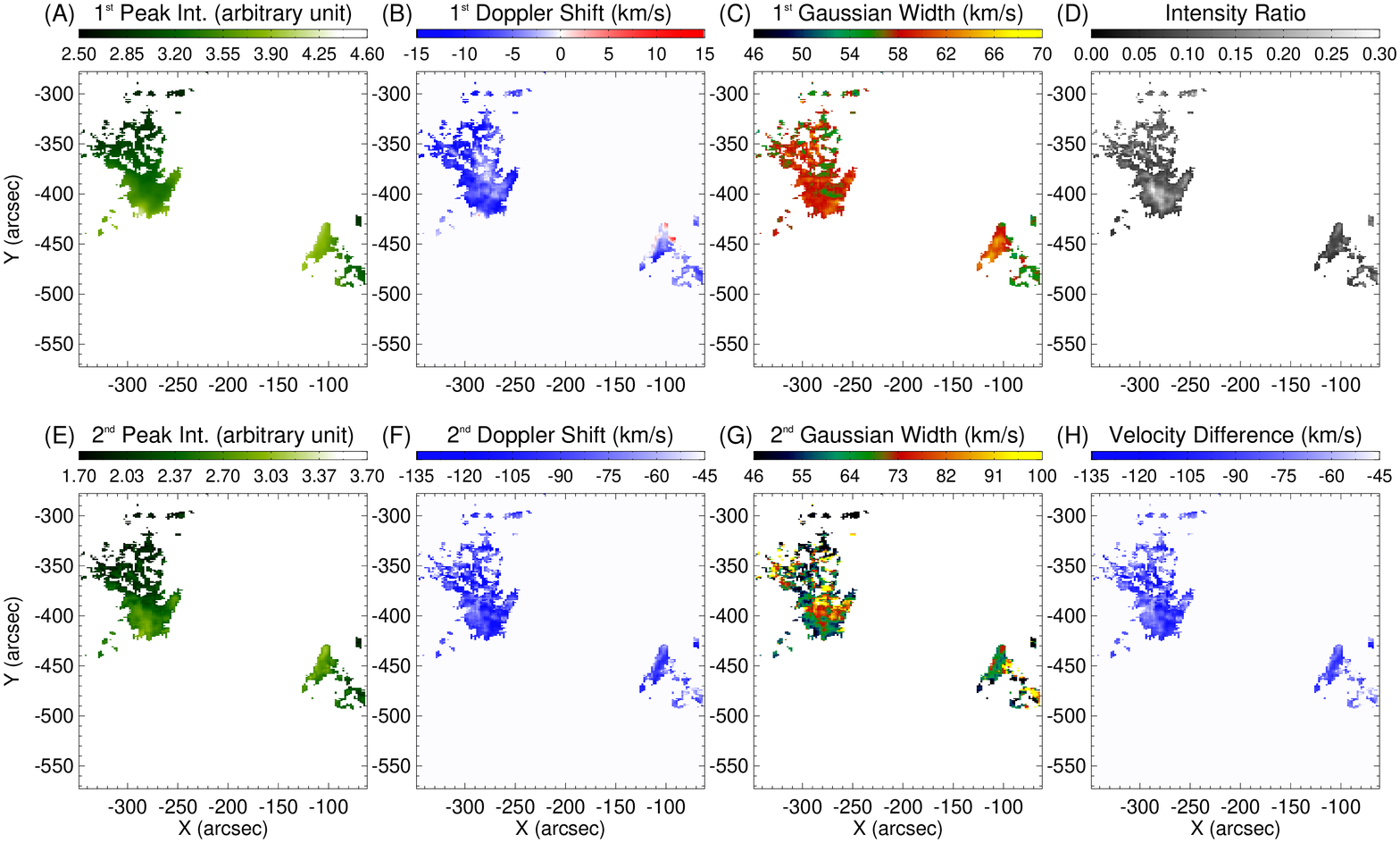}} \caption{Same as Figure~\pref{fig.5} but for the observation on 2010 September 16.}
\label{fig.10}
\end{figure*}

The RB$_{P}$ guided double Gaussian fit and RB$_{D}$ analysis were then applied to all spectra of Fe~{\sc{xiv}}~274.20\AA{} with obvious blueward asymmetries. Figure~\pref{fig.10} shows the spatial distributions of the three Gaussian parameters of both components as well as the intensity ratio and velocity difference
of the two components. Again, we found that the primary component is usually slightly blue shifted by $\sim$10~km~s$^{-1}$ in this observation. The velocity of the secondary component is blue shifted by much larger values. The maps of parameters derived from RB$_{D}$ asymmetry profiles are very similar to those derived from RB$_{P}$, which have been presented in Figure~\pref{fig.7} and thus are not shown here. Structures on the maps of RB$_{P}$ peak relative intensity, velocity, and width shown in Figure~\pref{fig.7} generally coincide with those on the maps of intensity ratio, velocity difference, and secondary component width as presented in Figure~\pref{fig.10}, indicating that both the RB$_{P}$ technique and RB$_{P}$ guided double Gaussian fit are able to extract the parameters of the secondary component.

\begin{figure}
\centering {\includegraphics[width=0.48\textwidth]{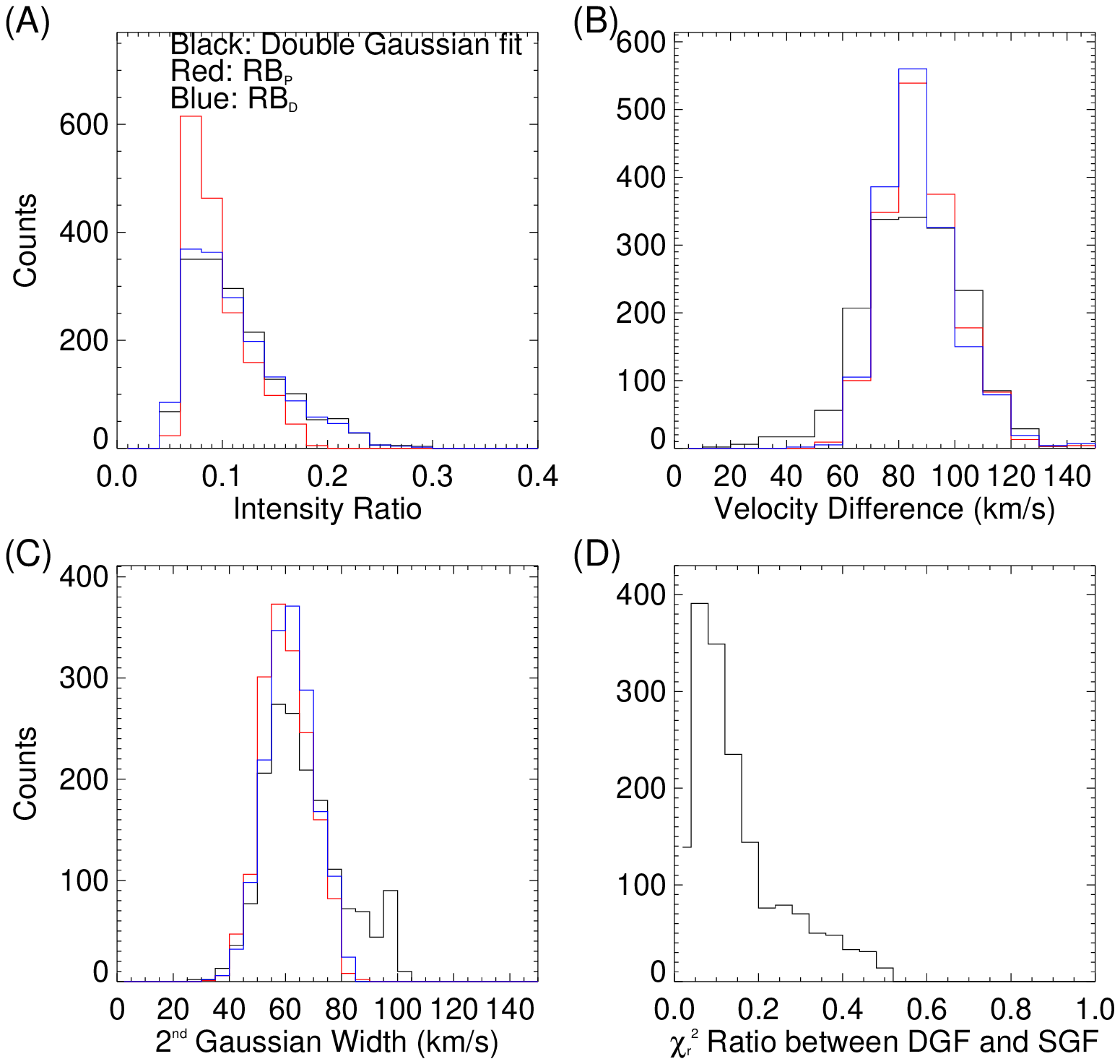}} \caption{Same as Figure~\pref{fig.6} but for the observation on 2010 September 16.}
\label{fig.11}
\end{figure}

Figure~\pref{fig.11}(A)-(C) shows the distributions of the intensity ratio (relative intensity), velocity difference (velocity), and Gaussian width of the secondary component as derived from double Gaussian fit and RB asymmetry analysis (RB$_{P}$,RB$_{D}$). For each of the parameters, the three histograms are very similar. All these histograms show no big difference from those presented in Figure~\pref{fig.6}, except the slightly larger peak velocity and width. The velocity of the secondary component is usually in the range of 50-120~km~s$^{-1}$ in this observation, which is about twice as high as that suggested by \cite{Peter2010}.

The distribution of the $\chi_{r}^{2}$ ratio between the double and single Gaussian fit in Figure~\pref{fig.11}(D) is also very similar to that in Figure~\pref{fig.6}(D). The ratio is usually less than 0.5, indicating a significant improvement of the fitting by using the double Gaussian fit instead of the single Gaussian fit. 

\subsection{AIA observations}

\begin{figure*}
\centering {\includegraphics[width=0.98\textwidth]{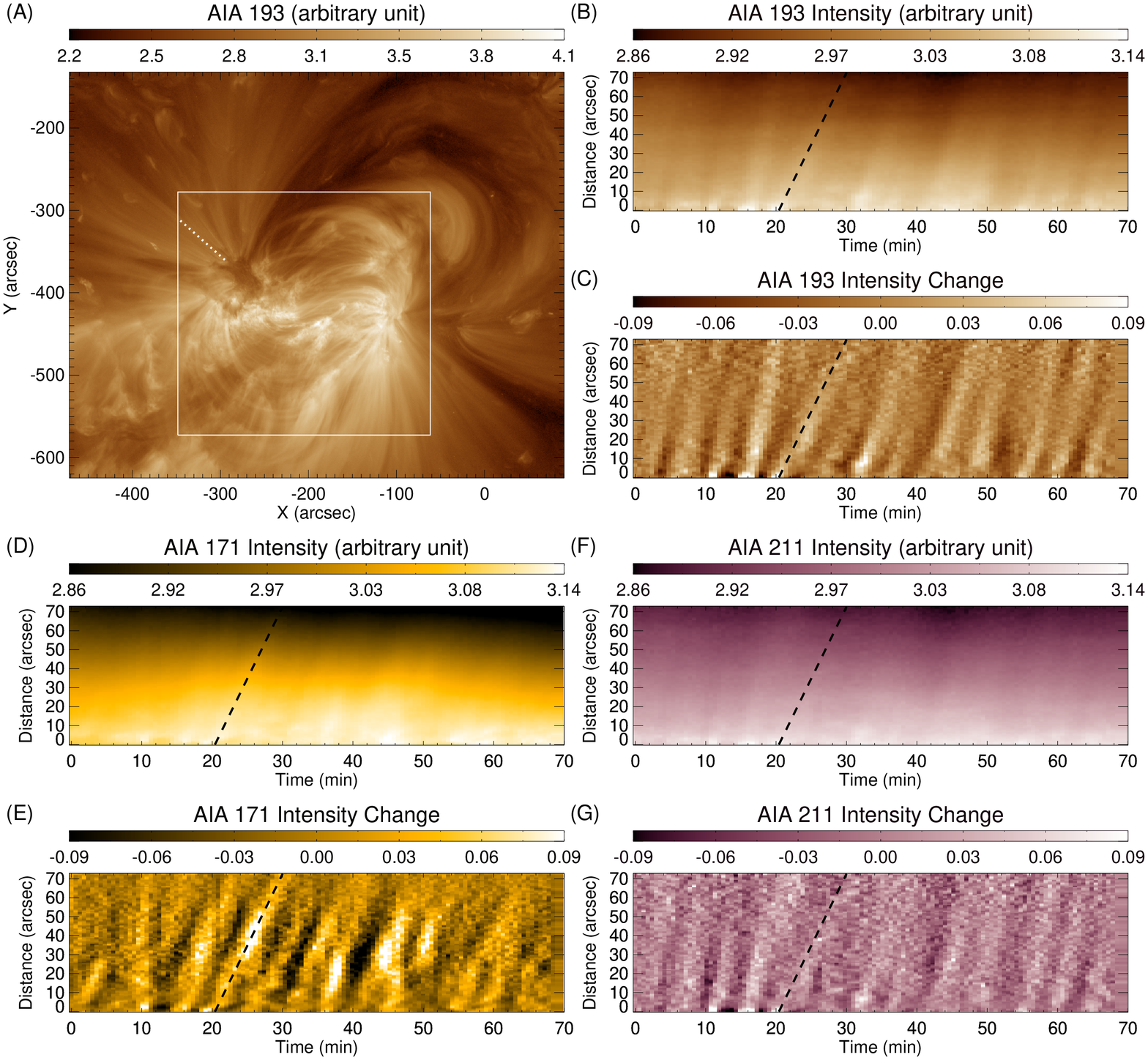}} \caption{(A) The image of AIA 193\AA{} at 12:06 on 2010 September 16. The rectangle
outlines the region scanned by EIS. (B) A space-time plot of the AIA 193\AA{} intensity for the dotted line shown in (A). (C) Same as (B) but showing the
detrended intensity. (D) \& (E): Same as (B) \& (C) but for AIA 171\AA{}. (F) \& (G): Same as (B) \& (C) but for AIA 211\AA{}. The inclined dashed line indicates one propagating disturbance with a traveling speed of about 91~km~s$^{-1}$. A movie associated with (A) is available on line (AIA193.mpeg). }
\label{fig.12}
\end{figure*}

During the EIS scan, the AIA instrument took full-disk images of the Sun in several EUV channels at a cadence of 12~s. We selected images in the passbands of 171\AA{}, 193\AA{} and 211\AA{} for our analysis. The dominant emission of these passbands has a temperature around 0.8 MK, 1.3 MK, and 2 MK, respectively. There were a few missing frames in each passband so that an linear interpolation was applied to the data. The pixel size is about 0.6$^{\prime\prime}$.

We extracted a sub-region enclosing the AR observed by EIS and co-aligned different image frames. A movie of AIA 193\AA{} images and the running
difference images is available online (AIA193.mpeg). Figure~\pref{fig.12}(A) shows an image of AIA 193\AA{} at 12:06. The rectangle indicates the region scanned by EIS. The cross correlation technique was used to coalign AIA 193\AA{} and EIS Fe~{\sc{xiv}}~274.20\AA{}.

Upward propagating disturbances (PDs) are clearly present in the AIA movie. These disturbances are mainly found at the two boundaries of the AR and coincide with obvious blueward asymmetries as seen from Figure~\pref{fig.7}(H), suggesting a connection of the two. The speed of
these disturbances can be estimated by placing a virtual slit along the propagation direction of the disturbances and calculating the slope of a
bright strip in the space-time (S-T) plot \citep[e.g.,][]{Sakao2007,McIntosh2010a,Tian2011b}. Figure~\ref{fig.12}(B)-(G) show the S-T plots of the original and detrended intensities for the dotted line shown in Figure~\pref{fig.12}(A). The detrended intensities were used to better reveal the faint outflow signatures and they were obtained by first subtracting a 8-minute running average from the original intensity time series and then normalized to the running average at each location of the slit. One example of PD is indicated by the inclined dashed line. The speed of this PD, the slope of the dashed line, is estimated to be 91~km~s$^{-1}$. 

As the PDs are usually present at different passbands of AIA, in principle we should be able to compare the speed of a certain
PD at different temperatures using AIA data. For a certain virtual slit, we often find similar patterns of the S-T plots
(both the original and detrended intensities) in the 193\AA{} passband and the other two passbands (particularly the 211\AA{} passband). In most cases we did not find an obvious increasing trend in the speed with increasing temperature, which seems to support the interpretation of PDs as multi-thermal high-speed upflows \citep[e.g.,][]{Sakao2007, McIntosh2009a,DePontieu2010,Tian2011a} rather than slow mode waves \citep[e.g.,][]{DeMoortel2000,DeMoortel2002,Robbrecht2001,King2003,McEwan2006,Marsh2009,Marsh2011,Wang2009a,Wang2009b,Verwichte2010,Stenborg2011}. However, a recent investigation suggests that non-dominant cool ions such as O~{\sc{v}} may contribute significantly to the total emission of the 193\AA{} and 211\AA{} passbands \citep{Martnez-Sykora2011b}. But cool ions such as O V would not be expected to show obvious emission over a large height range. For example, PDs are also seen off-limb in a coronal hole up to heights of 50-100 Mm with AIA  \citep{Tian2011b}, i.e., heights where emission from cool lines does not typically occur at sufficient strength. This renders the impact of non-dominant cool ions less likely. However, we can not rule out the possibility of slow waves. Perhaps both slow waves and upflows are existing in our observations. In fact waves can be exited by high-speed flows. But our EIS result that lines with different formation temperatures basically show similar velocities seems to suggest that upflows dominate the emission. We note that for some virtual slits the S-T plots show some differences in different AIA passbands, especially between the 171 \AA{} passband and the other two passbands. The emission in the 171 \AA{} passband is dominated by Fe~{\sc{ix}}~171.107\AA{} \citep{ODwyer2010}, which is formed in the upper transition region and thus may have a smaller scale height compared to the hotter emission in the other two passbands. Moreover, due to the different sensitivity, signal to noise ratio, and emission contrast in different passbands we may not always be able to clearly identify every outflow event in all the three passbands from the S-T plots. The 211 \AA{} passband samples plasma with a temperature much closer to the formation temperature of Fe~{\sc{xiv}}~274.20\AA{}. However, the 211 \AA{} images have a lower signal to noise ratio and are thus noisier compared to the 193 \AA{} images. Since the S-T plots in the 211 \AA{} passband are highly similar to those in the 193 \AA{} passband, in the following we concentrate on the 193\AA{} images.

\begin{figure}
\centering {\includegraphics[width=0.48\textwidth]{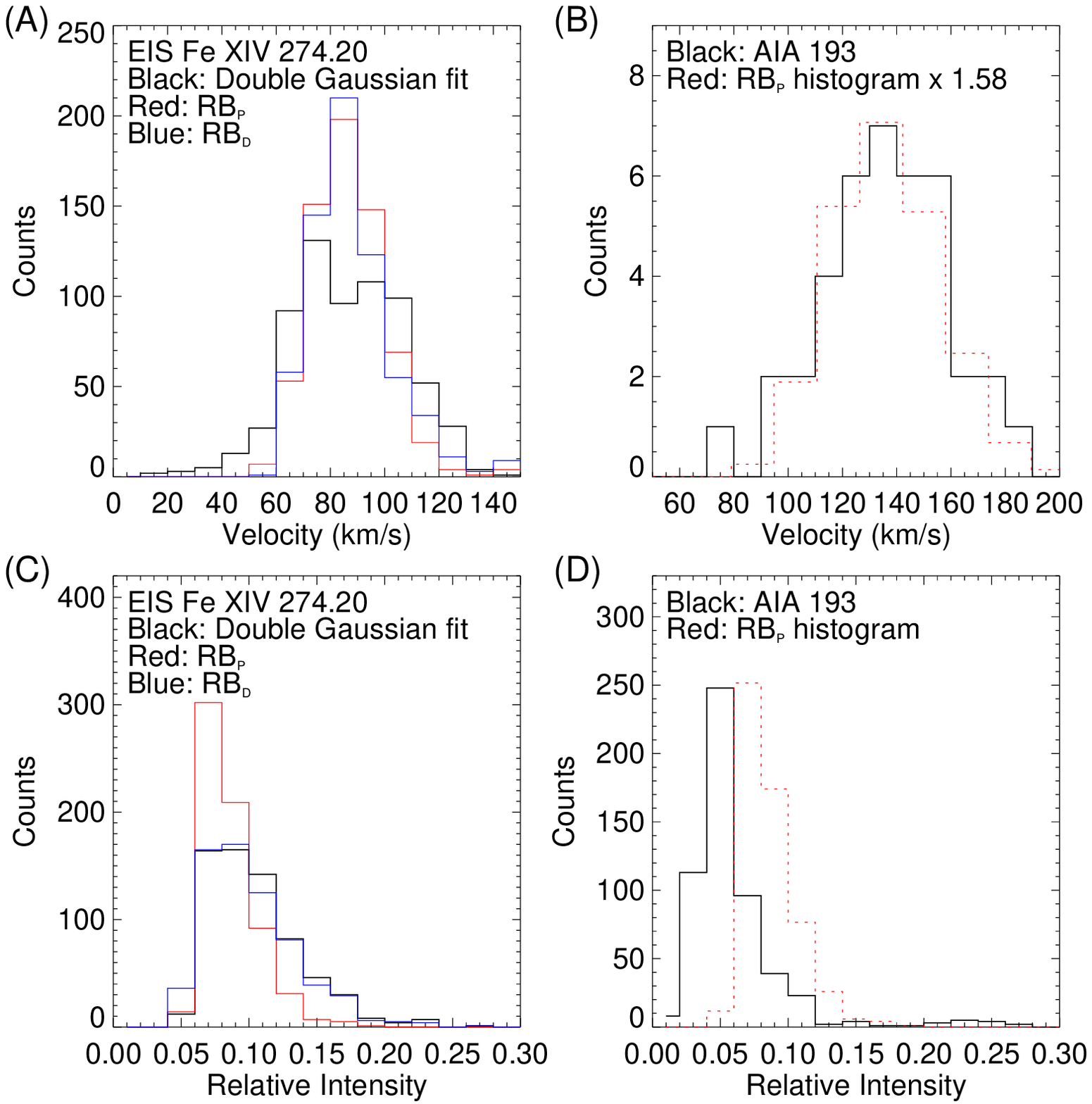}} \caption{Histograms of the velocity and relative intensity derived from the EIS Fe~{\sc{xiv}}~274.20\AA{} line profiles (A, C) and AIA 193\AA{} S-T plots (B, D) in the fan-like region at the left boundary of NOAA AR 11106. In (A) \& (C) the black, red and blue histograms represent results of the double Gaussian fit, RB$_{P}$, and RB$_{D}$ analysis, respectively. For comparison, the RB$_{P}$ velocity histogram is multiplied by 1.58 and overplotted as the dotted line in (B), and the RB$_{P}$ relative intensity histogram is overplotted in (D) as the dotted line. } \label{fig.13}
\end{figure}

We identified 39 well-isolated disturbances propagating along the fan-like structure at the left boundary of the AR and calculated their propagating speeds in
the 193\AA{} passband. A speed distribution was then obtained. A comparison of this speed distribution with the distribution of the velocities
derived from EIS Fe~{\sc{xiii}}~202.04\AA{} line profiles (through double Gaussian fit, RB$_{P}$, and RB$_{D}$ analysis) is presented in Figure~\ref{fig.13}(A) \& (B). Here only line profiles with obvious blueward asymmetries in the fan-like structure at the left boundary of the AR were used. By using the Potential Field Source Surface (PFSS) model \citep{Schatten1969,Schrijver2003}, the inclination angles of the magnetic field lines in the fan-like structure at the left boundary of the AR were estimated to be around 55$^{\circ}$ with respect to the line of sight. Disturbances propagating at a speed of $V$ along these magnetic field lines should exhibit a plane-of-sky speed of $V$$sin55^{\circ}$ and a line-of-sight speed of $V$$cos55^{\circ}$. The ratio of the two speed components is 1.43. Since the speeds derived from S-T plots of AIA images are the real speeds projected onto the plane of sky and the speeds derived from EIS line profiles
are the line-of-sight component of the real speeds, the ratio of the two should be around 1.43. From Figure~\ref{fig.13} we find that this is
indeed the case. The speeds derived by using AIA data are systematically larger than those derived by using EIS data. The ratio of the two
average speeds is around 135/85=1.58, which is close to 1.43. For comparison, the RB$_{P}$ velocity histogram is multiplied by 1.58 and overplotted as the dotted line in Figure~\ref{fig.13}(B). The small difference between 1.58 and 1.43 could easily be caused by the fact that the PFSS model is not expected to give a perfect representation of what viewing angles the real Sun presents. We think that in the future 3-D reconstruction \citep[e.g.,][]{Feng2007,Aschwanden2008} could be considered as an alternate method to help relate the velocities derived from imaging and spectroscopic observations. In Figure~\ref{fig.13} we can also compare histograms of the relative intensities derived from EIS Fe~{\sc{xiii}}~202.04\AA{} line profiles (in panel (C), through double Gaussian fit, RB$_{P}$, and RB$_{D}$ analysis) and those derived from S-T plots (D). The latter was obtained through the following process: we first smoothed the original time series with a 1-hour kernel. After that we subtracted the smoothed time series from the original one and obtained a time series of intensity change. The relative intensity was then obtained by normalising the maximum intensity change to the average intensity at each slit location. From Figure~\ref{fig.13}(D) we can see that the relative intensity is usually a few percent but can also be larger than 10\%, which is consistent with magnitude of the relative intensity of the secondary component revealed by RB asymmetry analysis and double Gaussian fit. For a better comparison, the RB$_{P}$ relative intensity histogram is overplotted as the dotted line in Figure~\ref{fig.13}(D). Note that the lack of small values of the relative intensity derived from EIS line profiles are due to the fact that we only selected profiles with significant blueward asymmetries. These results provide further support to the argument that the blueward asymmetries of coronal line profiles are caused by the upward propagating disturbances at AR boundaries \citep{McIntosh2009a,Tian2011a}. 

As mentioned by \cite{DeMoortel2009}, although there are striking similarities, it is not clear how the outflows inferred from EIS line profiles and the PDs seen in EUV imaging observations are related. \cite{DeMoortel2009} pointed out that the blue shifts (from single Gaussian fit) of EIS lines are only of the order of 20-50 km s$^{-1}$ and no periodicity has been reported. While the PDs in imaging observations have a speed of $\sim$100~km~s$^{-1}$ and they are often quasi-periodic. Now we understand that the EIS line profile consists of two components and that the secondary component has a velocity of $\sim$100~km~s$^{-1}$. And we have already demonstrated that the secondary component is also quasi-periodically enhanced or weakened \citep{DePontieu2010,Tian2011a}. The consistency between properties of the PDs observed by AIA and those of the secondary component revealed by RB asymmetry analysis and double Gaussian fit of EIS line profiles suggests a close connection of the two. It is thus natural to suggest that the PDs are responsible for the blueward asymmetries of line profiles. The non-thermal width of the secondary component is probably caused by both the superposition of multiple unresolved upflows and the Alfvenic motions associated with the upflows \citep{McIntosh2009a,Dolla2011}.  

We have to mention that in Figure~\ref{fig.13} we only present the results for the fan-like structure at the northeast boundary of the AR because of the simple magnetic field structure there. From the online movie (AIA193.mpeg) we can see that outward propagating disturbances are also clearly present in the fan-like structure to the southwest of the AR. We produced several S-T plots and found generally similar S-T patterns in different passbands. However, the footpoints of these fan loops (around x=$-110^{\prime\prime}$, y=$-450^{\prime\prime}$) are mixed together and the higher parts of different loops are not well-separated, making it difficult to identify a well-isolated propagating disturbance in the S-T plots. In other parts of the AR, magnetic field structures are either too complex or only slightly inclined with respect to the line of sight so that reliable speeds may not be obtained through S-T plots.

\section{Conclusion}
In conclusion, the presence of asymmetric coronal line profiles in loop footpoint regions suggests that at least two emission components are present: an almost stationary primary component and a high-speed secondary component. The secondary emission component might play an important role in supplying the corona with hot plasma. 

We have generated artificial spectra composed of two Gaussian components to test the ability of extracting the Gaussian parameters
(intensity, velocity, and width) of the secondary component through the originally defined (use the single Gaussian fit to determine the line
centroid) and a slightly modified (use the spectral position corresponding to the peak intensity as the line centroid) technique of the RB
asymmetry analysis, and find that this technique, especially the modified one, can provide a relatively accurate estimate of the parameters of
the secondary component if the velocity is larger and the width is not considerably larger than the primary component width. We have applied the
modified technique to the spectra obtained by the EUV Imaging Spectrometer (EIS) onboard Hinode in two active regions and found that the
distributions of the velocity/width determined from the RB asymmetry analysis peak around 80 km s$^{-1}$/60 km s$^{-1}$.  A comparison of such
distributions with the values of velocity/width calculated from the artificial spectra suggests that the secondary component cannot be
very broad or narrow and that the widths of the two components are comparable. The velocity of the secondary component is usually within the range of 50-150~km~s$^{-1}$ and seldom reaches as high as 200~km~s$^{-1}$ or as low as 40~km~s$^{-1}$ in our EIS observations. The relative intensity of the secondary component with respect to the primary one is often a few percent but can sometimes reach 30\%. Our conclusions are very different from those of \cite{Peter2010}, who claimed that the secondary component
contributes some 10\% to 20\% to the total emission, is about twice as broad as the primary component and blueshifted by up to 50~km~s$^{-1}$. The velocity range of the secondary component we derived here is roughly consistent with, yet slightly smaller than that derived by \cite{Bryans2010}. However, we find that the very high speed of 200~km~s$^{-1}$ mentioned by \cite{Bryans2010} seems to be not very likely. 

Using the parameters determined from the RB asymmetry
analysis and single Gaussian fit as initial values for several free parameters, we have performed a double Gaussian fit to the spectra that were observed to have obvious blueward asymmetries. The RB asymmetry analysis has been performed for the third time, but then using the centroid of the secondary component determined from the double Gaussian fit as the line centroid. The double Gaussian fit and the RB asymmetry analyses have given basically consistent results. The properties of the secondary component stay relatively stable across the temperature range of log({\it T}/K)=5.8-6.3. The double Gaussian fit also shows that the velocity of the primary component is often blue shifted by $\sim$10~km~s$^{-1}$. 

We have also used imaging observation simultaneously performed by the AIA instrument onboard {\it SDO} and demonstrated that the propagating disturbances coincide with obvious blueward asymmetries of line profiles, show an intensity change similar to the relative intensity of the secondary component of line profiles and an average velocity consistent with that derived from emission line profiles. Such results suggest that the upward propagating disturbances are plasma upflows responsible for the blueward asymmetries of line profiles.

\begin{acknowledgements}
{\it SDO} is the first mission of NASA$^{\prime}$s Living With a Star (LWS) Program. EIS is an instrument onboard {\it Hinode}, a Japanese
mission developed and launched by ISAS/JAXA, with NAOJ as domestic partner and NASA and STFC (UK) as international partners. It is operated by
these agencies in cooperation with ESA and NSC (Norway). Scott W. McIntosh is supported by NASA (NNX08AL22G, NNX08BA99G) and NSF (ATM-0541567, ATM-0925177). Bart De Pontieu is supported by NASA grants NNX08AL22G and NNX08BA99G. Hui Tian is supported by the ASP Postdoctoral Fellowship Program of NCAR. The National Center for Atmospheric Research is sponsored by the National Science Foundation. Hui Tian thanks Marc DeRosa for his assistance in using the PFSS package. We appreciate the efforts of the referee to improve the manuscript.
\end{acknowledgements}

\begin{appendix}
\section{A comparison between RB$_{P}$ and RB$_{S}$ results}
\begin{figure*}
\centering {\includegraphics[width=0.98\textwidth]{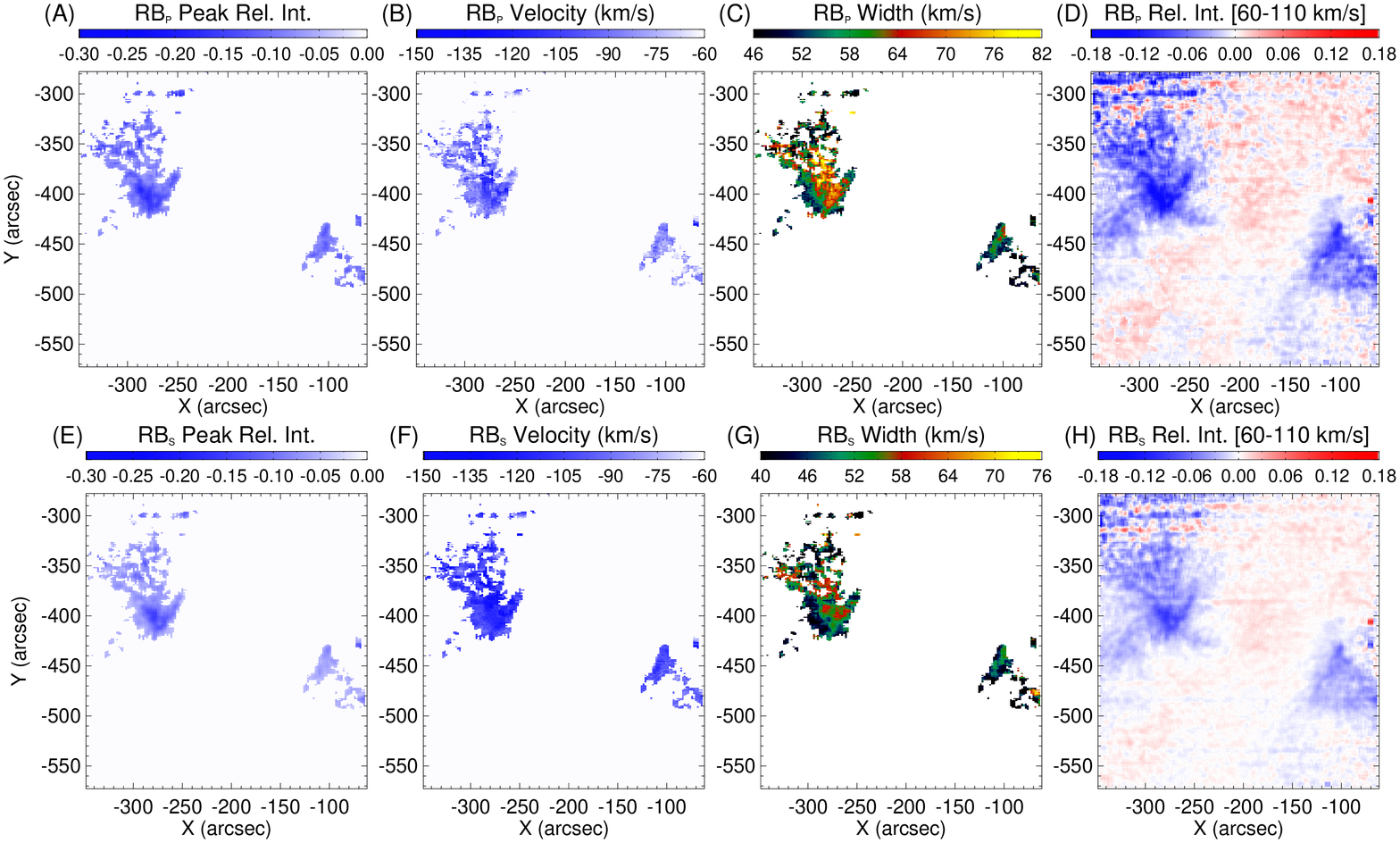}} \caption{Spatial distributions of the parameters derived from RB$_{P}$ (A-D) and RB$_{S}$ (E-H) asymmetry analysis for Fe~{\sc{xiv}}~274.20\AA{} in the observation on 2010 September 16. } \label{fig.s1}
\end{figure*}

As we mentioned in Section 2, both the RB$_{P}$ and RB$_{S}$ techniques can provide a relatively accurate estimate of the Gaussian parameters of the secondary component when the offset velocity is larger than the Gaussian width of the primary component. We have also found that the modified RB technique, RB$_{P}$, has a better ability to accurately quantify the properties of the secondary component as compared to the originally defined RB$_{S}$ technique. Figure~\ref{fig.s1} shows a comparison between the RB$_{P}$ and RB$_{S}$ results for the 2010 September 16 observation. We can see that the spatial structures in all of the four parameters are very similar. The most prominent difference is the smaller values of relative intensity and width, and larger upflow velocity for RB$_{S}$. Such results are fully consistent with our forward modeling results in Figure~\ref{fig.1} and imply that in our previous work the relative intensity and width might be underestimated while the upflow velocity was likely to be slightly overestimated.

\section{A comparison between EIS Fe~{\sc{xiv}}~274.20\AA{} and Fe~{\sc{xiv}}~264.78\AA{} line moments and profile asymmetries}
\begin{figure*}
\centering {\includegraphics[width=0.98\textwidth]{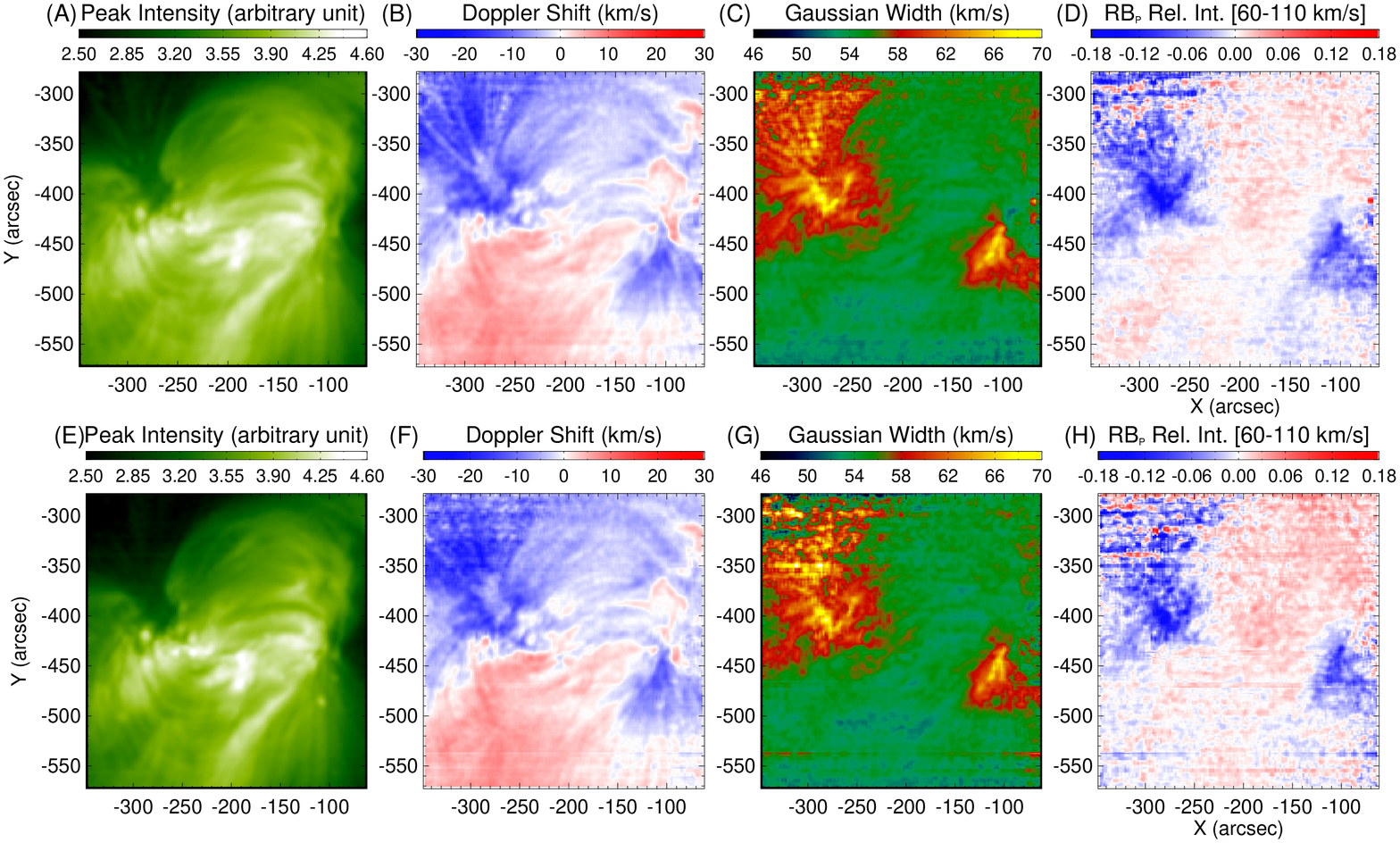}} \caption{Spatial distributions of the line moments (single Gaussian parameters) and profile asymmetries (average relative intensity in the velocity interval of 60-110~km~s$^{-1}$, as obtained from the RB$_{P}$ profiles) for Fe~{\sc{xiv}}~274.20\AA{} (A-D) and Fe~{\sc{xiv}}~264.78\AA{} (E-H) in the observation on 2010 September 16. } \label{fig.s2}
\end{figure*}

We have mentioned that the strong Fe~{\sc{xiv}}~274.20\AA{} line is blended with the weak Si~{\sc{vii}}~274.18\AA{} line and that the latter can safely be ignored in active region conditions \citep{Young2007}. The very close wavelengths (less than 1 spectral pixel) of the two also suggest that the blend is very unlikely to have an important impact on the high-speed (of the order of 100~km~s$^{-1}$) upflow. The Fe~{\sc{xiv}}~264.78\AA{} line spectra were not obtained in the 2007 January 18 observation so that we can not provide a direct comparison of it with the Fe~{\sc{xiv}}~274.20\AA{} line. In 2010 September 16 both Fe~{\sc{xiv}} lines were used in the observation and Figure~\ref{fig.s2} shows the spatial distributions of the line moments and profile asymmetries. Note that the spectra (especially for the Fe~{\sc{xiv}}~264.78\AA{} line) in the upper left corner, the higher part of the fan, have low signal to noise ratio and thus the line moments and profile asymmetries are noisy there. The very similar behavior of line moments and profile asymmetries between Fe~{\sc{xiv}}~274.20\AA{} and the weaker but clean Fe~{\sc{xiv}}~264.78\AA{} line provides further confidence that the blend of Si~{\sc{vii}}~274.18\AA{} line can be safely ignored in our analysis. 

\section{EIS observation on 2007 December 12}

We also applied the techniques of RB asymmetry analysis and double Gaussian fit to an EIS observation of AR NOAA 10978 from 11:43 to 17:03 on 2007 December 12. \cite{Bryans2010} found that the secondary emission component in this AR has velocities around 110~km~s$^{-1}$ and, as they stated in their paper, often as high as 200~km~s$^{-1}$. \cite{Bryans2010} used the Fe~{\sc{xii}}~195.12\AA{} line and they mentioned that the blend of this line, Fe~{\sc{xii}}~195.18\AA{}, should not significantly affect the Fe~{\sc{xii}}~195.12\AA{} line in the low-density \citep{Doschek2008,Brooks2011} outflow region (AR edges). However, to avoid the complexity introduced by this blend we decided to turn to other clean and strong Fe~{\sc{xii}} lines. We found that in the AR edges some other lines at the far blue wing of the Fe~{\sc{xii}} 192.39\AA{} line can be relatively strong and thus affect the results of RB asymmetry analysis and double Gaussian fit. So here we mainly present the results of the Fe~{\sc{xii}} 193.51\AA{} line, which is clean and about 2/3 as strong as Fe~{\sc{xii}}~195.12\AA{}. Since the line is very strong, there is no need to perform a spatial average of the line profiles.

\begin{figure*}
\centering {\includegraphics[width=0.98\textwidth]{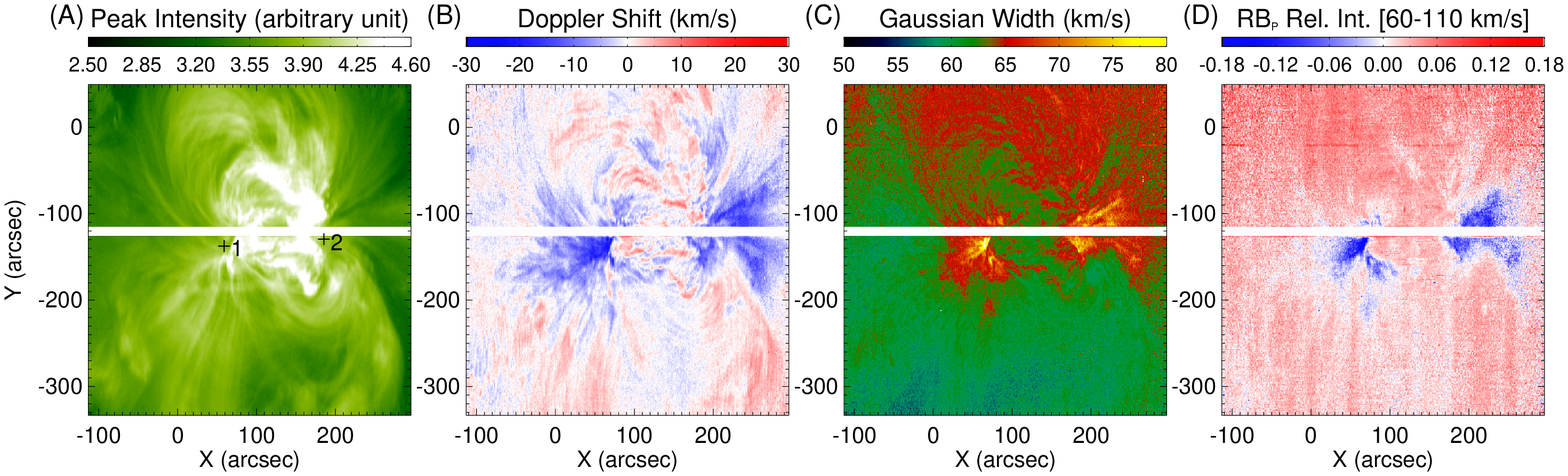}} \caption{Spatial distributions of single Gaussian parameters and profile asymmetries (average relative intensity in the velocity interval of 60-110~km~s$^{-1}$, as obtained from the RB$_{P}$ profiles) for Fe~{\sc{xii}}~193.51\AA{} in the observation on 2007 December 12. The two pluses mark locations of the line profiles presented in Figure~\ref{fig.s4}(A)\&(B), respectively.} \label{fig.s3}
\end{figure*}

Figure~\ref{fig.s3} shows the spatial distributions of the single Gaussian parameters and profile asymmetry. Similar to AR 10938 and AR 11106, the two loop footpoint regions are characterized by an enhancement in the blue shift, line width, and blueward profile asymmetry, and there is an obvious correlation between the Doppler shift/Gaussian width and profile asymmetry. Note that the white strip in each panel mark locations where the observed data is lost.

\begin{figure*}
\centering {\includegraphics[width=0.98\textwidth]{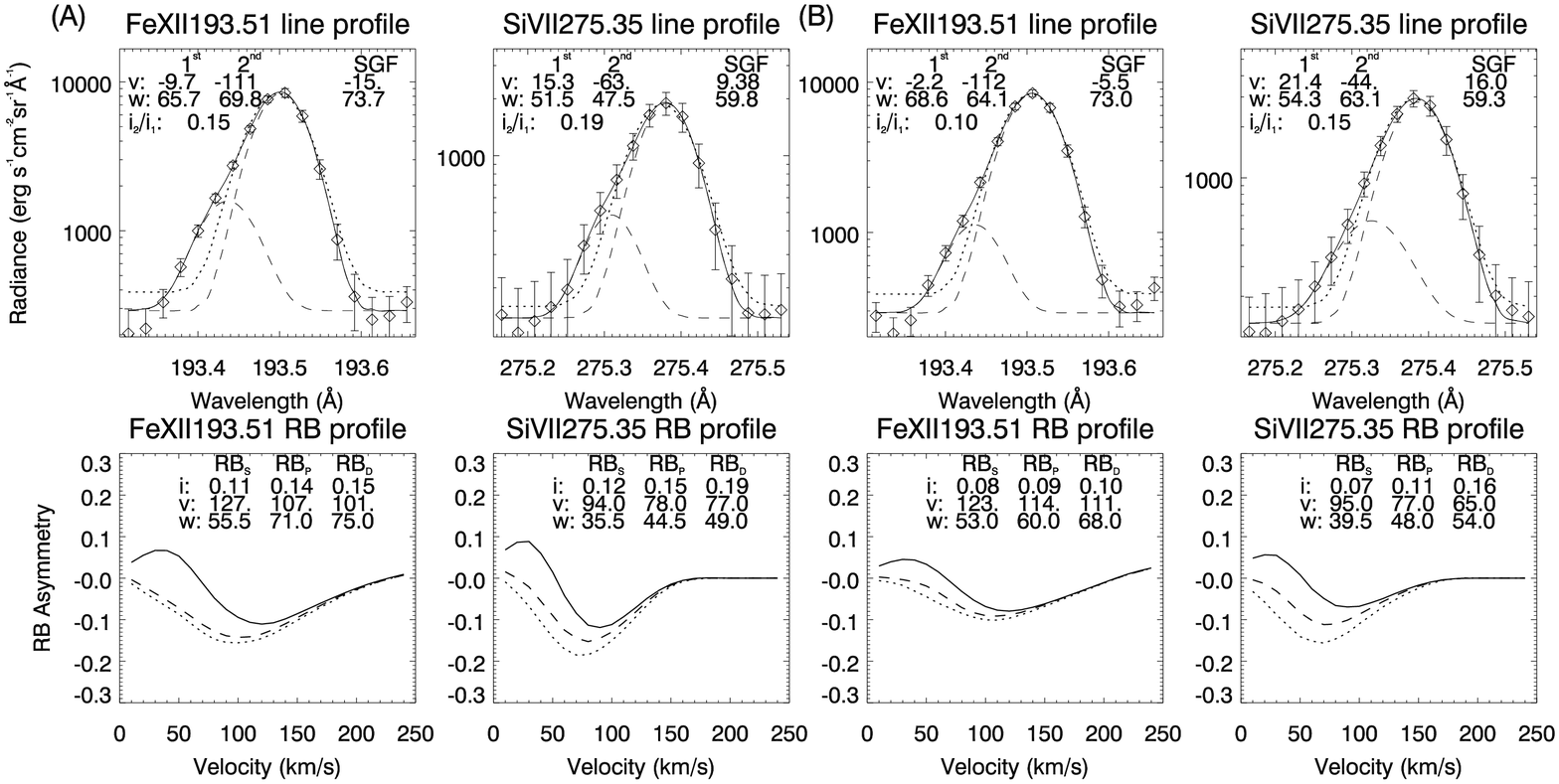}} \caption{ (A) RB asymmetry profiles of the Fe~{\sc{xii}}~193.51\AA{} and
Si~{\sc{vii}}~275.35\AA{} line profiles at location 1 marked in Figure~\pref{fig.s3}(A). The line styles and denotations of parameters are the same as in Figure~\pref{fig.4}. (B) Same as (A) but for the line profiles at location 2 marked in Figure~\pref{fig.s3}(A).} \label{fig.s4}
\end{figure*}

Two examples of line profiles and the corresponding RB asymmetry profiles of the Fe~{\sc{xii}}~193.51\AA{} and
Si~{\sc{vii}}~275.35\AA{} lines are shown in Figure~\pref{fig.s4}. The profiles were averaged over 3 pixels centered at the locations marked in Figure~\pref{fig.s3}(A) in both spatial dimensions. \cite{Bryans2010} mentioned that cool lines like Si~{\sc{vii}}~275.35\AA{} do not have any significant asymmetries and that their profiles can be accurately represented by a single Gaussian. We found that this is not always the case. Similar to what we found in AR 11106, at some locations the Si~{\sc{vii}}~275.35\AA{} line profiles clearly exhibit obvious asymmetries. As shown in Figure~\pref{fig.s4}, usually the primary component is redshifted by $\sim$15~km~s$^{-1}$ and the velocity difference between the two components is smaller than that of Fe~{\sc{xii}}~193.51\AA{}, which might be due to the complexity introduced by cooling downflows.  

\begin{figure*}
\centering {\includegraphics[width=0.98\textwidth]{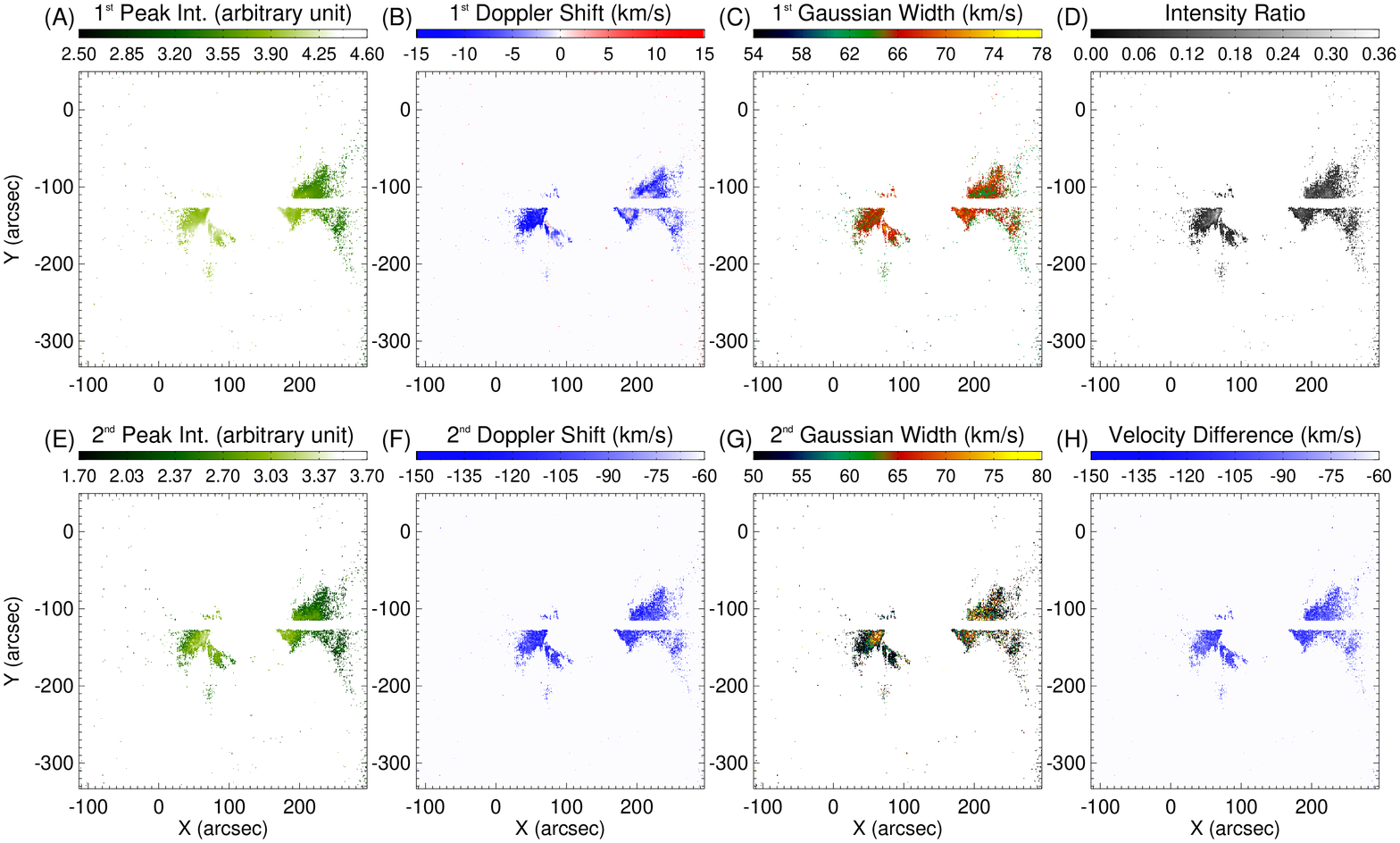}} \caption{Same as Figure~\pref{fig.5} but for the observation on 2007 December 12.}
\label{fig.s5}
\end{figure*}

Similar to what we did for AR 10938 and AR 11106, the techniques of RB asymmetry analysis and RB$_{P}$ guided double Gaussian fit were applied to each spectral profile with an obvious blueward asymmetry (the average RB asymmetry in 60-110~km~s$^{-1}$ smaller than -0.05). Spatial distributions of the double Gaussian parameters are presented in Figure~\pref{fig.s5}. Spatial distributions of the Gaussian parameters derived from the RB$_{P}$ and RB$_{D}$ analyses are similar to those of the secondary component shown in Figure~\pref{fig.s5}, and thus are not shown here. 

\begin{figure}
\centering {\includegraphics[width=0.48\textwidth]{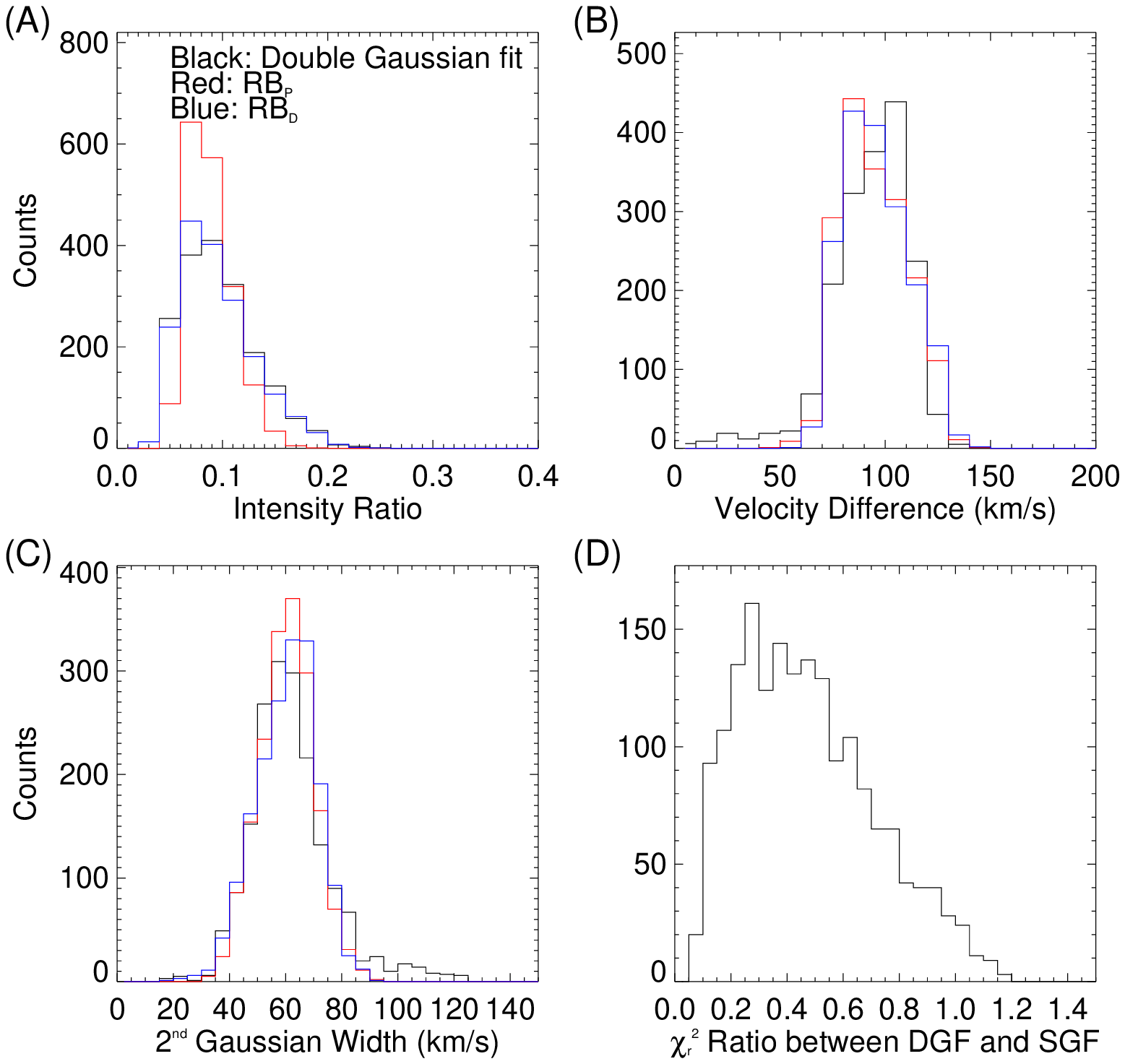}} \caption{Same as Figure~\pref{fig.6} but for the observation on 2007 December 12.}
\label{fig.s6}
\end{figure}

The distributions of the intensity ratio (relative intensity), velocity difference (velocity), and Gaussian width of the secondary component as derived from double Gaussian fit and RB asymmetry analysis (RB$_{P}$,RB$_{D}$), as shown in Figure~\pref{fig.s6}(A)-(C), reveal no big difference from those in Figure~\pref{fig.11}. However, the velocity distributions seem to peak at $\sim$95~km~s$^{-1}$, a value larger than those in AR 10938 ($\sim$75~km~s$^{-1}$) and AR 11106 ($\sim$85~km~s$^{-1}$) and close to that derived by \cite{Bryans2010} ($\sim$110~km~s$^{-1}$). The velocity seldom reaches as high as 200~km~s$^{-1}$. In fact from Figure 7 of \cite{Bryans2010} one can conclude that the velocity of the secondary component is usually in the range of 60-170~km~s$^{-1}$ and that only a very minor portion of the velocities exceeds 200~km~s$^{-1}$.

As seen from Figure~\pref{fig.s6}(D), the $\chi_{r}^{2}$ ratio between the double and single Gaussian fit is generally smaller than 1, indicating an improvement of the fitting by using the double Gaussian fit instead of the single Gaussian fit.

\end{appendix}

\end{document}